\documentclass[utf8]{frontiersSCNS} 
\usepackage{url,hyperref,lineno,microtype,subcaption}
\usepackage[onehalfspacing]{setspace}
\def\apss{Astroph.   Space Sci.}
\def\mgii{Mg{\sc ii}$\lambda$2800\/}
\def\apj{Astroph. J.}
\def\aj{Astron. J.}
\def\apjl{Astroph. J. Lett.}\def\apjs{Astroph. J. Supp.}
\def\mnras{Mon. Not. R. Astron. Soc.}
\def\aap{Astron. Astroph. }\def\pasp{Publ. Astron. Soc. Pacific}
\def\pasj{Publ. Astron. Soc. Japan}
\def\nat{Nature}
\defcitealias{marzianietal16}{M16}
\defcitealias{marzianietal16a}{M16a}
\def\mbh{$M_\mathrm{BH}$}

\def\nh{$n_\mathrm{H}$}
\def\kms{km s$^{-1}$\/}
\def\hb{H$\beta$}
\def\oii{[O{\sc ii}]$\lambda$3727\/}
\def\rfe{$R_\mathrm{FeII}$\/}
\def\lledd{$L/L_\mathrm{Edd}$}
\def\hbbc{H$\beta_\mathrm{BC}$}
\def\aliii{Al{\sc iii}$\lambda$1860\/}
\def\ciii{C{\sc iii}$\lambda$1909\/}
\def\siiii{Si{\sc iii}]$\lambda$1892\/}
\def\civ{C{\sc iv}$\lambda$1549\/}
\def\siiv{{Si{\sc iv}$\lambda$1397,1402\/}}
\def\siii{{Si{\sc ii}$\lambda$1816\/}}

\def\oiiiopt{[O{\sc iii}]$\lambda\lambda$4959,5007}
\def\feii{Fe{\sc ii}}
\def\civonly{C{\sc iv}}

\newcolumntype{C}[1]{>{\centering\let\newline\\\arraybackslash\hspace{0pt}}m{#1}}
\def\keyFont{\fontsize{8}{11}\helveticabold }
\def\firstAuthorLast{Marziani {et~al.}} 
\def\Authors{
P. Marziani\,$^{1,*}$,  D. Dultzin\,$^{2}$, J. W. Sulentic\,$^{3}$, A. Del Olmo\,$^{3}$,  C. A. Negrete\,$^{2}$, M. L. Mart\'{i}nez-Aldama\,$^{3}$,  M. D'Onofrio\,$^{4}$,  E. Bon\,$^{5}$, N. Bon\,$^{5}$, G. M. Stirpe\,$^{6}$}
\def\Address{$^{1}$ INAF, Osservatorio Astronomico di Padova, Italy \\
$^{2}$\ Instituto de Astronom\'{i}a, UNAM, Mexico\\
$^{3}$\ Instituto de Astrof\'{i}sica de Andaluc\'{i}a (IAA-CSIC), Granada, Spain\\ 
$^{4}$ Universit\`a di Padova, Padova, Italia\\  
$^{5}$Belgrade Astronomical Observatory, Belgrade, Serbia\\
$^{6}$INAF, Osservatorio di Astrofisica e Scienza dello Spazio di Bologna, Italia  }
\def\corrAuthor{Paola Marziani}
\def\corrEmail{paola.marziani@oapd.inaf.it}
\begin{document}
\onecolumn
\firstpage{1}
\title[ ]{A main sequence for quasars} 
\author[\firstAuthorLast ]{\Authors} 
\address{} 
\correspondance{} 
\extraAuth{}
\maketitle
\begin{abstract}
The last 25 years saw a major step forward in the analysis of optical and UV spectroscopic data of large quasar samples. Multivariate statistical approaches have led to the definition of systematic trends  in observational properties that are the basis of physical and dynamical modeling of quasar structure. We  discuss the empirical correlates of the so-called  ``main sequence'' associated with the quasar Eigenvector 1, its governing physical parameters and several implications on our view of the quasar structure, as well as some luminosity effects associated with the virialized component of the line emitting regions. We   also briefly discuss quasars in a segment of the main sequence that includes the strongest FeII emitters. These sources show a small dispersion around a well-defined Eddington ratio value, a property which makes them potential Eddington standard candles.
\tiny
 \keyFont{ \section{Keywords:}  galaxy evolution -- quasars -- eigenvector 1 -- outflows --  emission lines  --  supermassive black holes -- black hole physics} 
\end{abstract}

\section{Introduction}
\label{intro}

A defining property of   type-1 quasars is the presence of broad and narrow optical and UV lines emitted by ionic species over a wide range of ionization potentials \citep[IPs, ][]{vandenberketal01} which can be conveniently grouped in  high-ionization lines (HILs) involving IP $\gtrsim$ 50 eV, and low-ionization lines from ionic species with IP $\lesssim 20$ eV (Table 1). Optical and UV lines do not all show the same profiles, and quasar redshifts measured on different lines often show significant differences.   Internal line shifts ({ i.e.,  differences in redshift from different emission lines measured for the same object}) involve both broad and narrow emission lines  and   have offered a powerful diagnostic tool  of the quasar innermost structure and of the emitting region dynamics since a few years after the discovery of quasars \citep{burbidgeburbidge67}.  This is true even if the assumption that unobscured type-1 quasars have very similar properties has remained widespread until very recent times. This assumption has been, in our opinion, one of the most damaging prejudices in the development of quasar research. At a meeting in 1999 in Mexico City Deborah Dultzin half-jokingly suggested that ``a thousand spectra are worth more than one average spectrum'' as an extension of the  aphorism ``a spectrum is worth a thousand pictures'' \citep{dultzin-hacyanetal00}. The developments in the last 15+ years have proved that this is indeed the case, although the value of single-epoch observations may have gone under appreciated  with respect to other lines of evidence. We will therefore focus the scope of this review mainly to the organization of single-epoch spectra of large samples of quasars, following the ``bottom-up'' approach developed by Jack Sulentic and by his  collaborators.  

Over the years,  the UV resonance line CIV$\lambda$1549 { was considered  }as representative of broad  HILs, and  [OIII]$\lambda$4959,5007 were employed as   strong and easily accessible narrow HILs. Typical broad LILs include Balmer lines, FeII emission as well as MgII$\lambda$2800.\  { Their analysis requires an accurate measurement of the quasar redshift. Narrow LILs (\hb\ and \oii) have been found to be the best estimator of the quasar systemic redshift which defines the quasar ``rest-frame'' \citep[e.g.,][]{eracleoushalpern03,huetal08,bonetal18}.  The representative  narrow and broad HILs \oiiiopt\  and \civ\ show systematic blueshifts with respect to LILs in a large fraction of type-1 AGN (see e.g., \citealt{gaskell82,tytlerfan92,corbinboroson96,marzianietal96,richardsetal02}, and  \citealt{zamanovetal02,zhangetal11,marzianietal16,shenetal16}  for broad and narrow lines, respectively). Broad LILs could be used as a last resort especially in high-redshift quasars where the UV rest-frame is accessible from optical observations and no narrow lines are observable \citep{negreteetal14a}. Even if broad LILs can show significant shifts with respect to rest frame, these are infrequent, and rarely as large as those found among HILs.}

The interpretation of inter-line shifts in quasar spectra is mainly based on the Doppler effect due to gas motion with respect to the observer, along with selective obscuration. This explanation is almost universally accepted. For [OIII]$\lambda$4959,5007 blueshifts, it is consistent with a moderately dense outflow   ($\log n \sim 2 -  5 $ [cm$^{-3}$]) of  optically thin gas. For CIV$\lambda$1549, the explanation is not fully consistent if line emission occurs from optically thick clouds distributed, for example, symmetrically in a bicone whose axis is aligned with the spin  of the black hole \citep[e.g., ][]{zhengetal90}. Such optically thick outflows might more easily give rise to a net redshift, if the receding part of the outflow remains visible.  While it is no longer under discussion that large \civonly\ blueshifts (amplitude $\gtrsim 1000$ km/s) involve radial motion + obscuration, \citet{gaskellgoosmann13} suggested an alternative explanation involving infall and ``reflection.'' The observer does not  see photons  from the line emitting gas, but  photons backscattered toward  herself from a sea of hot electrons over the accretion disk. If the photons were originally emitted from infalling gas (i.e., approaching the disk), the observer should see a net blueshift.  This explanation has some appeal for CIV$\lambda$1549, but sounds very unlikely for [OIII]$\lambda$4959,5007 { because \oiiiopt\ is emitted on spatial scales ranging from a few pc to thousands of pc, where a suitable ``mirror'' as the one potentially offered by hot electrons surrounding the accretion disk may not exist}. In view of source  commonality in terms of CIV$\lambda$1549 and [OIII]$\lambda$4959,5007 blueshifts (\citealt[][]{marzianietal16}, \citealt{marzianietal16a}), we will { follow} the most widely accepted interpretation that blueshifts involve outflow and obscuration for both CIV$\lambda$1549 and [OIII]$\lambda$4959,5007. { The interpretation of LIL blue- and redshifts will follow the same assumption} (with some caveats,  \S \ref{acc}). 

{ This review will be focused on the way internal line shifts and other quasar properties can be efficiently organized. A major step was an  application to  quasar spectra of the Principal Component Analysis (PCA)    carried out in the early 1990s   \citep{borosongreen92,francisetal92}. \citet{borosongreen92} 
measured the most prominent emission features in the \hb\ spectral region, and found the first hint of the quasar ``main sequence'' (MS; their Fig. 9). The PCA and other statistical techniques require  measurable parameters for a set of sources. The starting point is therefore the definition of a set of parameters (Section \ref{sec:diag}) that may be conductive to the identification of fundamental correlations (the Eigenvectors) as well as to physics. The PCA of type-1 quasars (we remark in Section \ref{sec:unif} that we are dealing exclusively with type-1, unobscured quasars)  yields a first eigenvector from which the MS is defined (Section \ref{ms}). After reviewing the MS correlates (Section \ref{birdseye}), we show how the ``empirical'' eigenvector 1 can be connected to the main physical parameters of quasars seen as accreting systems (Sections \ref{sec:emp} and \ref{sec:accpar}).  The most intriguing results point toward two different accretion structures in type-1 quasars (Section \ref{sec:self}), which are however largely self-similar over a wide range in luminosity. As an example of the power of the MS to identify sources that are physically similar, we consider quasars located at the extreme tip of the MS which are potential distance indicators (Section \ref{sec:xa}).}

\begin{table}[!t]
\begin{center}
\vbox{\footnotesize\tabcolsep=17pt
\parbox[c]{204mm}{\baselineskip=12pt
{Table 1.}{ \sc Identification of high- and low-ionization lines }
\vspace{0.25cm}
}
\begin{tabular}{|c|c|c|c|c|} 
\hline
& Broad &  Narrow  &   low-$z$ & high-$z$ \\ \hline
HILs (IP $\gtrsim 50$ eV) & CIV$\lambda$1549, HeII$\lambda$1640 & [OIII]$\lambda\lambda$4959,5007, HeII$\lambda$1640,   &  Space & Visual \\
&NV$\lambda$1240 & [Ne III]$\lambda$3869  &&\\
\hline
LILs (IP $\lesssim 20$ eV) & HI Balmer (H$\beta$),  FeII,   & Balmer, [OII]$\lambda$3727,  [OI]$\lambda$6300 &  Visual & IR \\
& MgII$\lambda$2800, Ca IR Triplet & [SII]$\lambda\lambda$6716,6731 &&    \\
 \hline
\end{tabular}
}
\end{center}
\end{table}

\section{Diagnostics from single-epoch spectra: internal line shifts and intensity ratios}
\label{sec:diag}

Single-epoch spectroscopy of large quasar samples  yields  data  defined by limits in multiplexing ({ i.e., by the ability to obtain a record of signals in different wavebands with the same observations}): synoptical observations of the UV, visual, NIR rest frame have been challenging until a few years ago (and, in part, they are still challenging  to-date). Given the limit in multiplexing, observations of low- and high-$z$ quasars provide different information since different  rest-frame wavelength ranges are covered: visual spectrometers provide the rest-frame H$\beta$ range  at low $z$ but the rest-frame UV at $z \gtrsim 1.5$. To cover the rest-frame UV at low-$z$, space-based observations are needed. To cover H$\beta$ at high $z$, NIR spectroscopy is needed. These limitations are being overcome by new generation instruments mounted at the focus of 8m-class telescopes which provide simultaneous coverage of visual and NIR, but these facilities  were not available at the time most of the work reviewed in this paper was done. Synoptic observations of visual and UV at low-$z$ still require coordinated ground and space-based observations which are especially hard to come by. 

Table 1 provides an overview of the lines covered in the different domains. When we speak of intermediate to high-$z$\ objects (a population of behemoth quasars that is now extinguished),  we are speaking of sources that are not anymore observed at $z \lesssim 1$ { \citep[mainly for the ``downsizing'' of nuclear activity, see e.g.][and references therein]{springeletal05,sijackietal15,fraix-burnetetal17}.} On the other hand, at high-$z$, only relatively few quasars with luminosities  comparable to those of low-$z$\ quasars are known since their apparent magnitude would be too faint. They aren't yet efficiently sampled by the major optical source of quasar discovery, the Sloan Digital Sky Survey (SDSS, \citealt{blantonetal17}, and references therein). 

\subsection{Analysis}

A prerequisite for a meaningful analysis of internal line shifts and intensity is a spectral resolution $R = \lambda/\delta\lambda \gtrsim 1000$ and S/N$\gtrsim$20.  In addition,  the quasar rest frame must be known with good precision. As mentioned in the Introduction, accurate redshifts can be obtained by limiting the   measurements to narrow LILs (H$\beta$)  and [OII]$\lambda$3727, the latter with some caveats \citep{bonetal18}.  Narrow   HILs (e.g., [OIII]$\lambda\lambda$4959,5007)  show systematic blueshifts \citep{zamanovetal02,eracleoushalpern03,rodriguez-ardilaetal06,huetal08}  whose amplitude is a strong function of their location along the quasar main sequence (\S \ref{birdseye}).  

Once the rest frame is known, quantitative measurements of emission line profiles centroids, line widths at different fractional heights  become possible. Centroids  are defined by: $c(\frac{i}{4}) = (\mathrm{FW}(\frac{i}{4}) - 2 \cdot \lambda_{0})/2$, $\forall i=1,\ldots,4$, where the full width is FW$(\frac{i}{4}) = \lambda_{R}(\frac{i}{4}) - \lambda_{B}(\frac{i}{4})$. 

In more recent times, we have applied a heuristic multicomponent decomposition whose rationale will be given in \S \ref{multi}, that is in part equivalent to a inter-percentile profile analysis \citep{marzianietal10,shangetal07}. This more model-dependent approach has been used along with centroids and width measurements on the full profile.  Intensity ratios are computed separately for each profile component. The multicomponent fits have the advantage of isolating regions that are partly resolved in the radial velocity dimension and in different physical conditions (\S \ref{hillil}). 

\begin{figure}[htp!]
\begin{minipage}[t]{0.5\linewidth}
\centering
\includegraphics[width=3.1in]{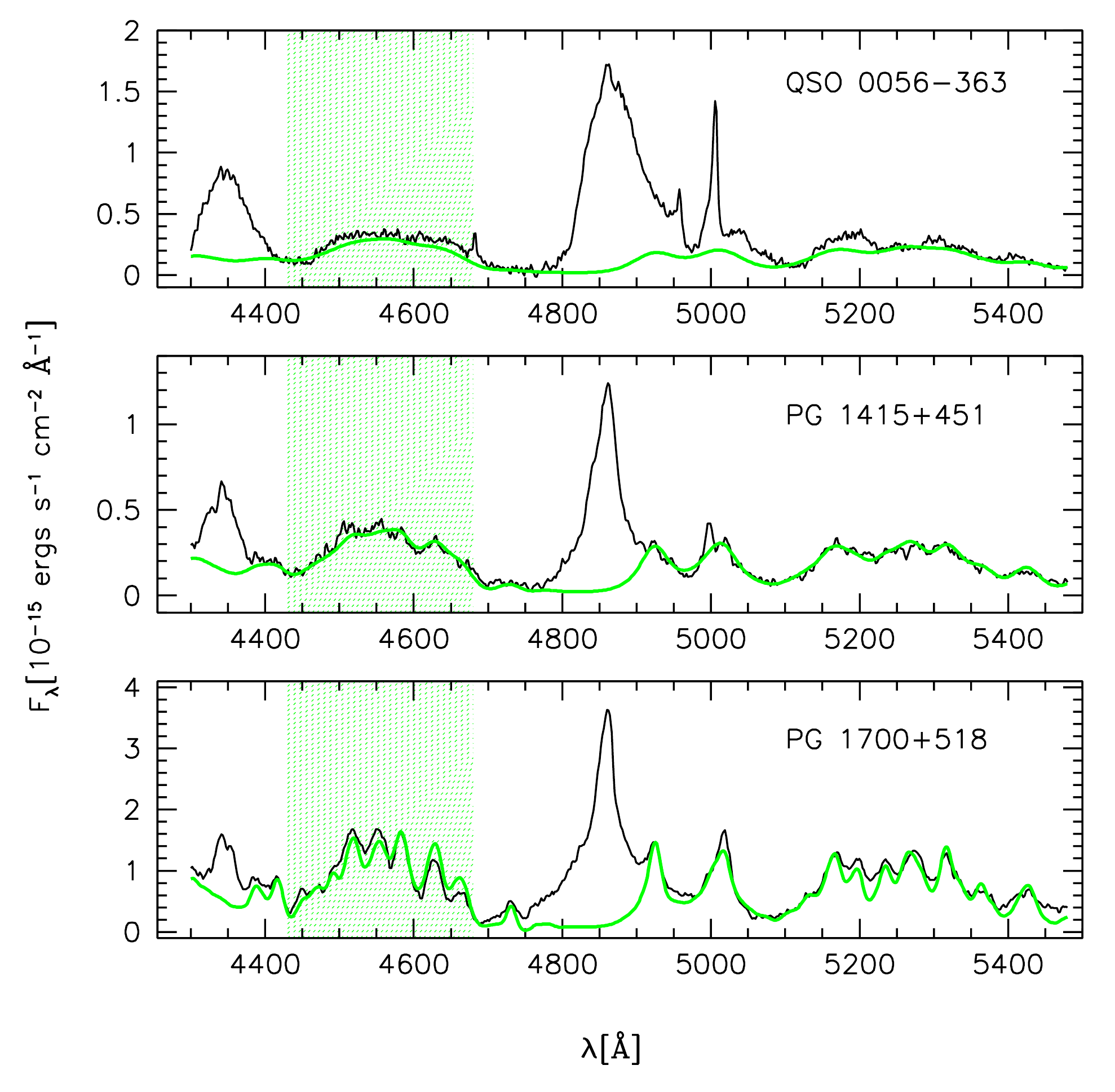}
\end{minipage}
\hspace{0.25cm}
\begin{minipage}[t]{0.45\linewidth}
\centering 
\vspace{-3.1in}
\caption{ Three type-1 quasars. The green line represents the FeII emission template, scaled and broadened to fit the observed FeII emission. The shaded area between 4340 and 4680 \AA\ covers the wavelength range used to compute  the total intensity of the blended emission at 4570 \AA\ to estimate the relative intensity between FeII $\lambda$ 4570 and \hb\ i.e., the \rfe\ parameter  ({ \citealt{borosongreen92}}, \S \ref{ms}). The three spectra show different \rfe: as the intensity ratio \rfe\ decreases { (from bottom to top)}, the \hb\ line width increases. Also notable is the change in profile shape of H$\beta$. The spectra of the three quasars are exemplary of general trends involving LILs observed along the quasar main sequence.}
\label{fig:figure2}
\end{minipage}
\end{figure}

An important element in the analysis of optical and UV lines is the measurement of the FeII emission contribution. As shown in Fig. 1 of \citet{marzianietal06}, FeII multiplets are strong in the optical and UV and even create a pseudo continuum  in the range 2100--3000 \AA. Contamination in the \hb\ spectral range is also strong. Within the limits of the past analysis, it has proven appropriate to assume that the FeII multiplet ratios are always the same, even if the FeII features change equivalent width and FWHM from object to object \citep{borosongreen92}. The trend shown in Fig. \ref{fig:figure2} motivates this assumption. In practice, it has been possible to model the FeII emission using a scaled and broadened template obtained from the Narrow Line Seyfert 1 (NLSy1) galaxy I Zw 1, a strong FeII emitter with narrow broad lines. More sophisticated approaches varying multiplet ratios \citep[e.g.,][]{kovacevicetal10} are needed in the rare cases where FeII emission appears peculiar. 

Given the quantitative measurements on line profile and line profile components, much of the past work has been inspired by  the Baconian approach \citep{bacon1902},  deriving inferences   by induction from observations without (much) prior benefit of  theory. In practice this translated into (1) classifying  data in a systematic way, to avoid a mixup of sources which are empirically different; (2) applying a quantitative but phenomenological description of the data, (3) performing uni- and multivariate statistics with a quantitative treatment of errors (including, for example, also the analysis of censored data).  Model inferences have been deduced from the data separating aspects that were strongly constrained (for example, physical conditions derived from nebular physics), from those that required more speculative assumptions. 


\section{Unification models and type-1 quasars}
\label{sec:unif}

Unification schemes  have provided a powerful conceptual framework suitable for  organizing the analysis of AGN. The precursor distinction between type 1 and 2 Seyfert \citep{khachikianweedman74} gained  a convincing interpretation when \citet{antonuccimiller85} reported the discovery of a broad line component visible in the polarized spectrum of Seyfert 2 nuclei  but invisible in natural light: the broad line region is hidden from view and only photons scattered by hot electrons toward our line of sight are received by the observer. This explanation remains alive and widely accepted today \citep[e.g.,][]{eunetal17}, even if we now know that is only a part of the story: type-2 AGN  differ for environmental properties \citep{dultzin-hacyanetal99,koulouridisetal06,villarroelkorn14}, may intrinsically lack  a BLR \citep{laor00} at very low accretion rates, or may even be unobscured normal type-1 under special  conditions \citep{marinuccietal12}. The point  here is that unification models of RQ AGN separate { two types of quasars on the basis of the viewing angle between the line of sight and the symmetry axis of the system (i.e., the spin axis of the black hole or the angular momentum vector of the inner accretion disk) but make no prediction on unobscured type 1 AGN.} Orientation effects are expected also for unobscured type-1s, as we should observe them in the range of viewing angles,  $0 \lesssim \theta \lesssim 45-60$. There is little doubt that broad line width is affected by orientation, especially for RL sources \citep[e.g.,][]{willsbrowne86,rokakietal03,sulenticetal03,jarvismclure06,runnoeetal14}: { a comparison between RL sources that are core-dominated (believed to be oriented with the jet axis close to the line-of-sight) and lobe-dominated (misaligned) shows that the \hb\ FWHM is larger in the latter class. This result  strongly suggests a flattened, axisymmetric structure for the BLR. }  For RQ objects, the evidence is not obvious and an estimate of $\theta$\ remains an unsolved problem at the time of writing.  However, orientation effects are certainly not enough to explain the diversity of quasar spectral properties.

\section{The quasar  main sequence}
\label{ms}


At  the time the \citet{borosongreen92} paper appeared, studies based on moderately sized samples (20-30 objects) were common and often reached confusing results from correlation analysis. The best example is the Baldwin effect  (an anti-correlation between   equivalent width of HILs and luminosity) which was found in some and then not found in similar samples without apparent explanation.\footnote{ The Baldwin effect was originally described by \cite{baldwinetal78} and later detected in several samples   \citep{laoretal95,willsetal99};   negative   results were concurrently obtained   \citep{wuetal83,wilkesetal99}. \citet{sulenticetal00a} discuss early works, and the main reason of this apparent contradiction.}  In this respect the sheer size of the \citet{borosongreen92} sample was a key improvement.  { A novel aspect was also the application of the PCA which considers each parameter as a dimension of a parameter space, and searches for a new parameter space with fewer dimensions (defined by linear combinations of the original parameters) as needed to explain most of the data variance  \citep[][]{murtaghheck87,marzianietal06}.}   The application of the  PCA  was not unprecedented in extragalactic astronomy \citep[e.g.][]{diazetal89} but was well suited to quasar data that appeared weakly correlated among themselves without providing   a clear insight of which correlations were the most relevant ones.  

The quasar Eigenvector 1 was originally defined  by a  PCA  of  $\approx$ 80 Palomar-Green (PG) quasars and associated with an anti-correlation  between strength of FeII$\lambda$4570, \rfe\ (or [OIII] 5007 peak intensity)  and   FWHM of H$\beta$ \citep{borosongreen92}. The parameter \rfe\ is defined as  the ratio between  the integrated flux of FeII$\lambda$4570 blend of multiplets,  and that  of the  \hb\  broad component:\footnote{The term broad component without the suffix BC is used here to identify the full broad profile excluding the \hb\ narrow component. In more recent times, we have distinguished between two components, the broad component  \hbbc\ and the very broad component  \hb$_\mathrm{VBC}$. { The \hbbc\ is associated with the  core of the \hb\ broad line, and the \ \hb$_\mathrm{VBC}$ with its broader base (see Section \ref{multi}).}} \rfe = I(FeII$\lambda$4570)/I(H$\beta$).  Since 1992, various aspects of  the Eigenvector 1 (E1) of quasars  involving widely different datasets as well multi-frequency parameters have been discussed in more than 400 papers,  as found on NASA ADS in August 2017
\citep{borosongreen92,sulenticetal00a,sulenticetal00b,sulenticetal07,dultzin-hacyanetal97,shangetal03,tangetal12, kuraszkiewiczetal09,maoetal09,grupe04}.  Earlier analyses have been more recently confirmed by the exploitation of  SDSS-based samples \citep{yipetal04,wangetal06,zamfiretal08,richardsetal11,kruzceketal11,marzianietal13,shenho14,sunshen15,brothertonetal15}.

The second eigenvector -- Eigenvector 2 -- was found to be proportional to luminosity, and eventually associated with the HIL Baldwin effect(s) that are the most-widely discussed luminosity effects in quasar samples    \citep{baldwinetal78,dietrichetal02,bianetal12}. The smaller fraction of variance carried by the Eigenvector 2  indicates that luminosity is not the major driver of quasar diversity, especially if samples are restricted to low-$z$. We will not further consider  HIL luminosity effects\footnote{HIL luminosity effects are subject to strong biases. It is as yet unclear whether such biases can entirely account for the weak luminosity effects observed in large samples.}  but only discuss the influence of luminosity on the LIL FWHM (\S \ref{highl}).

 
 

\begin{figure}[h!]
\begin{center}
\includegraphics[width=15.5cm]{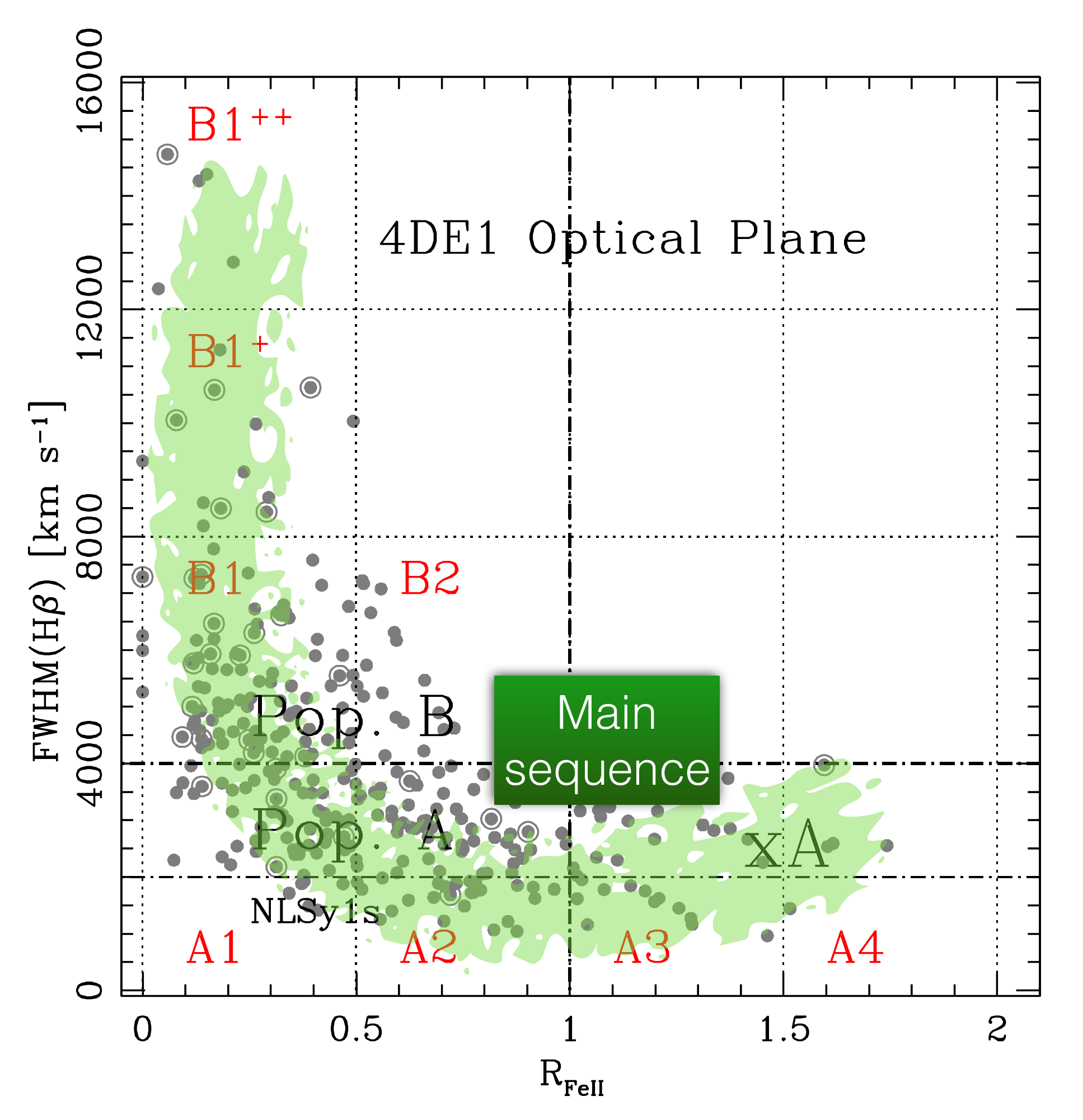}
\end{center}
\caption{The optical plane of the Eigenvector 1, FWHM(\hb) vs. \rfe. The green shaded area indicatively traces the distribution of a quasar sample from \citet{zamfiretal10}, defining the quasar MS. The thick horizontal dot-dashed line separates populations A and B; the thin identifies the limit of NLSy1s. The vertical dot-dashed line marks the limit for extreme Population  A (xA) sources with \rfe$\gtrsim 1$. Dotted lines separate spectral types, identified as explained in \S \ref{ms}.}
\label{figms} 
\end{figure}

The distribution of data points in the optical plane of the Eigenvector 1 FWHM(\hb) vs. \rfe\ traces the quasar main sequence (MS, Fig. \ref{figms}), { defined for quasars of luminosity $\log L \lesssim 47$ [erg s$^{-1}$], and   $z < 0.7$}. The MS shape allows for the definition of a sequence of spectral types (Fig. \ref{figms}), and  motivates subdividing  the 4DE1 optical plane into a grid of bins of FWHM(\hb) and FeII emission strength. Bins A1, A2, A3, A4 are defined in terms of increasing \rfe\ with bin size $\Delta$\rfe = 0.5, while bins B1, B1+, B1++, etc. are defined in terms of increasing FWHM(\hb) with $\Delta $FWHM = 4000 \kms. Sources belonging to the same spectral type show similar spectroscopic measurements (e.g., line profiles and line flux ratios). Spectral types are assumed to isolate sources with similar broad line physics and geometry. Systematic changes are reduced within each spectral type. If so, an additional advantage is that an individual quasar can be taken as a bona fide representative of all sources within a  spectral type. The binning adopted  (see Fig. \ref{figms}) has been derived for low-$z$ ($<0.7$) quasars.  Systematic changes may not be  eliminated in full, if an interpretation scheme such as the one of \citet{marzianietal01} applies, { who posited a continuous effect of Eddington ratio and viewing angle (at a fixed \mbh) as the origin of the MS shape (Section \ref{ulm} provides further explanations).}  


Developments in the analysis before late 1999 of low-$z$ quasar spectral properties are reviewed in \citet{sulenticetal00a}.  Data and ideas were in place as early as in year 2000 to introduce the idea of two quasar populations, A and B: Population A with FWHM(H$\beta$) $\le$ 4000 km s$^{-1}$; Population B (broader) with FWHM(H$\beta$) $>$ 4000 km s$^{-1}$ \citep{sulenticetal00a,sulenticetal00b}.  Later developments   have confirmed that the two populations are two distinct quasar classes. Population A may be seen as the class that includes local  NLSy1s as well as high accretors \citep{marzianisulentic14}, and Population B as { a class capable} of { high-power} radio-loudness \citep[][see Section \ref{birdseye}]{zamfiretal08}. It now seems  unlikely that the two populations are just the opposite extremes of a single quasar ``main sequence'' defined in the plane FWHM(\hb) vs. \rfe\ \citep[][\S \ref{birdseye}]{sulenticetal11}, { although the subdivision at FWHM(H$\beta$) $=$ 4000 km s$^{-1}$\ is not widely considered in literature. It is therefore worth analyzing the issue in some more detail after considering the main correlates along the MS.} 

\begin{figure}[hpt!]
\begin{center}
\includegraphics[width=18cm]{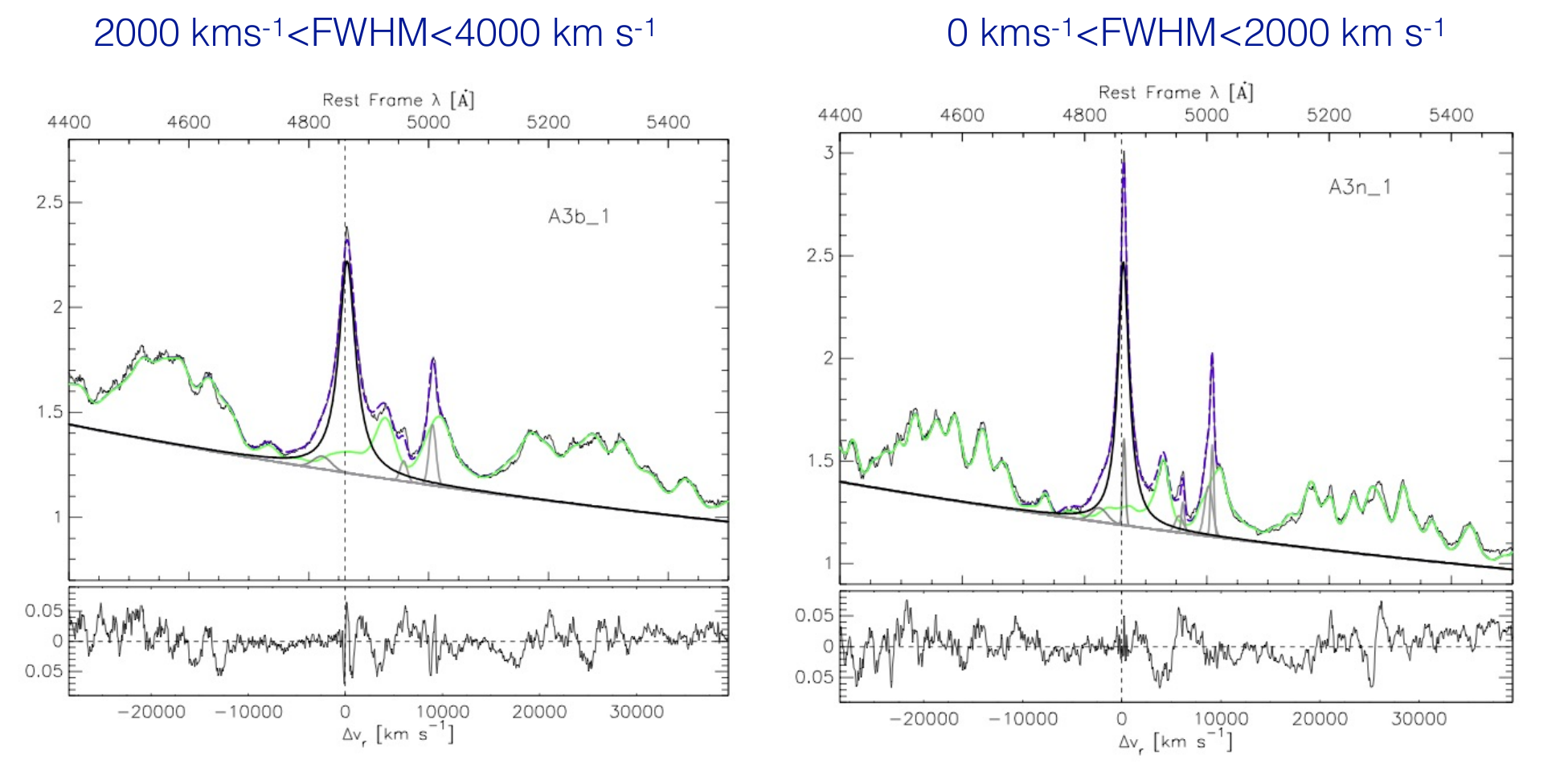} 
\end{center}
\caption{Fits of the A3 spectral type composites, obtained for the broader (A3b$\_$1) and narrow (A3n$\_$1) half of the spectral bin. The original spectrum (thin black line) is shown with FeII emission (pale green) superimposed to the continuum and  the \hb\ line with a Lorentzian profiles (thick black lines).  Thin grey lines trace \hb$_\mathrm{NC}$\ and \oiiiopt\ emission and, on the blue side of \hb, a faint excess emission that is not accounted for by the symmetric shape of \hb.  \label{fig:2}}
\end{figure}

\section{A bird's eye view of the MS correlates}
\label{birdseye}

Several correlates   have been proved as especially relevant in the definition of the MS multifrequency properties.
\begin{itemize}
\item { Balmer emission line profile shape --}  { Several past works found a clear distinction between Pop. A and B} in terms of Balmer line profile shapes { \citep{sulenticetal02,marzianietal03b}}: Pop. A sources  show Lorentzian Balmer line profiles, symmetric and unshifted; Pop. B, Double Gaussian (broad + very broad component, \hbbc\ + H$\beta_\mathrm{VBC}$, \S \ref{multi}), most often redward asymmetric. While several authors described the Balmer line profiles of NLSy1s as Lorentzian \citep[e.g.,][]{veroncettyetal01,craccoetal16}, the transition between the profile types is apparently occurring at 4000 km s$^{-1}$ and not at 2000 km s$^{-1}$, the canonical limit of NLSy1. This early result \citep{sulenticetal02} has been confirmed by several later analyses \citep[e.g.,][]{zamfiretal10,marzianietal13,negreteetal17}. Fig. \ref{fig:2} shows composite spectra in the FWHM range 2000 -- 4000 and 0 -- 2000 km s$^{-1}$: the profile shape remains the same as the line gets broader \citep{negreteetal17}. Mirroring  Paolo Padovani's prescription  as enunciated at the Padova meeting (\citealt{padovani17}: no more RL, only jetted!), we would recommend to speak of {Population A and B... and no more NLSy1s!} In both cases, it is not just a matter of nomenclature: inter-sample comparison will be biased if the subdivision is inappropriate.

\item { UV diagnostic ratios --} Major trends involve strong UV emission lines. Schematically, moving from spectral type B1++ to A4 we find:  NV$\lambda$1240/Ly$\alpha$: $\nearrow$; AlIII$\lambda$1860/SiIII]$\lambda$1892: $\nearrow$ CIII]$\lambda$1909/SiIII]$\lambda$1892 $\searrow$ W(NIII]1750) $\nearrow$ W(CIV$\lambda$1549) $\searrow$. These trends can be interpreted as an  increase in density and  metallicity and decrease in ionization parameter of the LIL-emitting part of the BLR toward the strongest FeII emitters at the tip of the MS \citep{baldwinetal96,willsetal99,bachevetal04,nagaoetal06,negreteetal12,negreteetal13}. 
\item { \civ\ centroid shifts } The CIV1549 centroid blueshifts are a strong function of a source location along the E1 MS, reaching maximum values in correspondence of the extreme Pop. A (xA, spectral types A3 and A4). They can be accounted for by a scaled, almost symmetric and unshifted profile (such as the one of H$\beta$)  plus an excess of blueshifted emission, corresponding to  a ``virialized'' emitting region plus an outflow/wind component, respectively \citep{marzianietal10}. The relative prominence of the two components is a function of the location on the MS: the outflow component can dominate in  xA sources, and be undetectable in sources at the other end of the MS  (B1++) where the broader profiles are found. If we measure the centroid shift at half maximum $c(\frac{1}{2})$, large blueshifts are found only in Pop. A  \citep{sulenticetal07}. The blueshifted excess is at the origin of a correlation between centroid shifts of \civ\ and  FWHM \civonly\ \citep{coatmanetal16,sulenticetal17}. This has important implications for \mbh\ estimates. 
\item { [OIII] blueshifts --}  The  [OIII]$\lambda\lambda$4959,5007 doublet mimics the blueshift  observed for \civ\ with respect to the rest frame.  
The average blueshift amplitude increases toward the high \rfe\ end of the MS \citep{zamanovetal02,marzianietal03c,zhangetal13,craccoetal16}. This is emphasized by the distribution in the OP of the [OIII]$\lambda\lambda$4959,5007 ``blue outliers'' (BOs)   which show  blueshift at peak of amplitude $\gtrsim$ 250 \kms. Large [OIII] shifts such as those of the BOs  are found for FWHM(H$\beta$)$<$ 4000 km/s. 
  \item { LIL blueshifts --} The profile of the resonance LIL   MgII$\lambda$2800  also suggests evidence of outflow (somewhat unexpectedly, \citealt{marzianietal13,marzianietal13a}): low ionization  outflows are detected in the  xA spectral types, but lower radial velocities are involved in MgII $\lambda$2800 than in CIV$\lambda$ 1549 ($\sim $ 100 vs. $\sim$ 1000 km/s).
\item { Radio loudness --} The probability of being RL is much larger among Pop. B sources: 25\%, among Pop. A  $\approx 3 - 4$\%\  \citep{zamfiretal08}.  Core-dominated RL sources are displaced toward Pop. A in the optical plane of the E1 because of orientation effects. \citet{zamfiretal08} suggest that RL sources should be considered as such only if very powerful with $\log P_{\nu} > 31.6$ [erg s$^{-1}$ Hz$^{-1}$], and Kellerman's $\log R_\mathrm{K} > 1.8$ \citep{kellermannetal89}, in line with the distinction of jetted and non-jetted suggested by \citet{padovani16} which considers as jetted only sources for which there is evidence of  powerful, relativistic  ejections. { On a broader perspective,   radio-loudness may not be restricted to low Eddington ratio, once the basic prescriptions  from the main mechanisms explaining jet formation and involving extraction of the rotational energy of the black hole or of the accretion disk in the presence of a large-scale, well-ordered, and powerful magnetic field are satisfied \citep[][]{blandfordznajek77,blandfordpayne82}.  Compact-steep spectrum (CSS) RL sources show  high radio power \citep{odea98} and  properties that are of Pop. A, with relatively high \lledd\ \citep{wu09b}. Formation of jetted  sources may occur also at high Eddington ratio, although the physical mechanism leading to jet production and collimation is presently unclear  \citep[for a review, see][]{czernyyou16}, and the jet properties may also be different \citep[][and references therein]{gu17}. If relatively low power is considered ($\log P_{\nu} \sim 31$ [erg s$^{-1}$ Hz$^{-1}$]), the RL Pop. A sources include RL NSLy1s \citep{komossaetal06}. The $\gamma$-ray detection for some of them \citep{abdoetal09,foschinietal10} may confirm  their ``jetted'' nature. It has been suggested that RL NLSy1s have CSSs as a mis-aligned parent population \citep{bertonetal16}.  Therefore, the absence of luminous RL sources among Pop. A sources may be related to the absence of highly-accreting very massive black holes \mbh ($\gtrsim 10^{9} M_{\odot}$) at relatively low-$z$ \citep[e.g.,][and references therein]{cavalierevittorini00,fraix-burnetetal17}. }
\item { Soft X-ray slope -- }  The steepness of the soft X-ray continuum measured by the photon index $\Gamma_\mathrm{soft}$ is also dependent on the location along the MS. $\Gamma_\mathrm{soft}$ is the measure of the soft-X excess (0.2 -- 2 KeV) above a canonical power law with $\Gamma \approx 2$. Values of $\Gamma_\mathrm{soft}$  $>$ 2 are mainly found for FWHM(H$\beta$) $<$ 4000 km/s \citep[i.e., in Pop. A, ][]{bolleretal96,wangetal96,sulenticetal00a,grupe04,shenho14,benschetal15}. 
\end{itemize}

Tables reporting main-sequence correlates are provided in several recent reviews and research papers \citep[e. g.][]{sulenticetal11,fraix-burnetetal17}, and in  Chapter 3 and 6 of \citet{donofrioetal12}. To restrict the attention of a subset of especially significant parameters, \citet{sulenticetal00a} introduced a 4D E1 parameter space. In addition to FWHM(\hb) and \rfe, two more observationally ``orthogonal'' parameters, $\Gamma_\mathrm{soft}$\ and $c(\frac{1}{2})$ \civ\	 are meant   to help establish a connection between observations and physical properties.  The  4DE1 parameters clearly support  the separation of Population A (FWHM \hb $<$4000 \kms) and Population B(broader) sources,\footnote{The 4000 \kms\ limit is appropriate at low redshift  and moderate luminosity $\log L \lesssim$ 47 [erg s$^{-1}$]; { see the discussion in Section \ref{highl}}.} although the non-optical parameters are not always useful since they are MS correlates and often unavailable.  The immediate interpretation of the 4DE1 parameters is summarized in Table 2. In the simplest term, the FWHM \hb\ is related to the velocity dispersion in the LIL emitting part of the BLR. On the converse, $c(\frac{1}{2})$ \civonly\ yields a measurement affected by the high-ionization outflow detected in the HIL profile. The largest $c(\frac{1}{2})$\ values indicate a decoupling between the strongest HIL and LIL features, with the latter remaining symmetric and unshifted \citep{marzianietal96}. The parameter \rfe\ is of more complex interpretation.  \rfe\ is  affected by the metallicity (obviously, if [Fe/H]=-10, \rfe$\approx$0) but metallicity is most likely not all of the story \citep{jolyetal08}, since \feii\ strength tends to saturate for high metallicity values. The main dependence is probably on ionization conditions, density and column density (\S \ref{ulm}). A $\Gamma_\mathrm{soft}> 2 $\ is usually ascribed to Compton thick  soft X-ray emission { from a hot corona above the disk, but may also be the high-energy tail of the spectral energy distribution of the disk itself, in   case} the inner disk is very hot \citep[e.g.,][]{doneetal12,wangetal14}.

\subsection{Pop. A and B: really a dichotomy?}

{ Supporting evidence in favor of a dichotomy between Pop. A and B includes the change in the \hb\ profile shape from Lorentzian-like to double Gaussian, with a redward asymmetry that is not detected in the narrower sources of Pop. A, not even in  spectral type A1 where \feii\ emission is weak. Large \civ\ centroid blueshifts are not observed in Pop. B unless sources of very high-luminosity are considered \citep{sulenticetal17,bischettietal17,bisognietal17}. Therefore, a dichotomy at $\approx$4000 \kms\  for $z \lesssim$ 1 and $\log L \lesssim $ 47 [erg s$^{-1}$] is empirically supported by a sudden change in observations parameters. On the other hand, if \rfe\ is mainly affected by \lledd\ and the \hb\ FWHM by the viewing angle, it is hard to justify a dichotomy   \citep[][]{shenho14}.  The analysis of \citet{marzianietal01} indicates that  spectral type A1 may include sources which are intrinsically of Pop. B and observed almost pole on. However, the FWHM \hb\ is also dependent on Eddington ratio \citep[][Fig. 3 of \citealt{marzianietal01}]{nicastro00}.  Most sources in bin A1 (\rfe $\le 0.5$) are true Pop. A with Lorentzian \hb\ profiles.  The Eddington ratio corresponding to the change in \hb\ and \civ\ properties  has been estimated to be $\approx 0.2 \pm 0.1$  in low-$z$, low-to-moderate $L$ samples \citep{sulenticetal00b,marzianietal03b}. This is also the limit for the presence of a fully thin accretion disk \citep[Section \ref{acc}; ][]{abramowiczetal88,szuszkiewiczetal96,abramowiczetal97}. At any rate, one should be aware that a fixed FWHM value only approximately corresponds to a well-defined Eddington ratio. The minimum FWHM for virialized systems (reached at maximum \lledd)  is luminosity-dependent (Sect. \ref{highl}), and so is the FWHM corresponding to any fixed \lledd\ even if the luminosity dependence is weak up lo $\log L \approx 47$ [erg s$^{-1}$] \citep[][]{marzianietal09}. }

\begin{table}[!t]
\begin{center}
\vbox{\footnotesize\tabcolsep=17pt
\parbox[c]{204mm}{\baselineskip=12pt
{Table 2 -- }{
\sc Interpretation of the 4FE1 parameter space  \label{tab lines}}
\vspace{0.25cm}
}
\begin{tabular}{|c|c|c|C{0.01cm}|c|c|}
\hline
Parameter &  Immediate interpretation  &  Relation to accretion  parameters  \\ 
&  &  and orientation at low-to-moderate $L$ ($\lesssim 10^{47}$ erg/s) \\ \hline
FWHM \hb & LIL-BLR velocity field   &  \lledd, $\theta$, \mbh\ \\
c(1/2) \civonly\ &  HIL-BLR velocity field (outflow) &  \lledd, $\theta$  \\
R$_\mathrm{FeII}$ &  LIL ionization and density, $Z/Z_{\odot}$ & \lledd, possibly $\theta$  \\ 
$\Gamma_\mathrm{soft}$ & Compton-thick X-ray emission, accretion disk emission & \lledd\\
 \hline
\end{tabular}
  }
\end{center}
\end{table}
  
\section{From empiricism to physics}
\label{sec:emp}

\subsection{Synergy between reverberation mapping and single-epoch spectroscopy of quasars}

Reverberation mapping { derived from the response of the emitting regions with respect to continuum changes} resolves the radial stratification of the line emitting regions and helps  assess their velocity field {\citep[e.g.,][]{peterson98,petersonhorne06}}.  The synergy with reverberation mapping helps to partially remove the spatial degeneracy of single-epoch spectra. 
The strongest evidence in support of virial motion for the line emitting gas comes from two lines of reverberation mapping investigation: (1) the anticorrelation between line broadening and the time lag of different lines in response to continuum variation \citep{petersonwandel99,petersonetal04}. This relation has been found for some  nearby Seyfert-1 nuclei with a slope close to the one expected from the virial relation. (2) Velocity-resolved reverberation mapping \citep[e.g., ][]{gaskell88,koratkargaskell91,grieretal13}  rules out outflow as the main broadening mechanism for LILs.  A general assumption adopted in the interpretation of single-epoch spectra is that  virial motion is indicated by almost symmetric and unshifted  line profiles (within $\approx$100 \kms\   from rest frame, the typical rest frame uncertainty for moderate dispersion spectra).   

\begin{figure}[htp!]
\begin{minipage}[t]{0.5\linewidth}
\centering
\includegraphics[width=8.5cm]{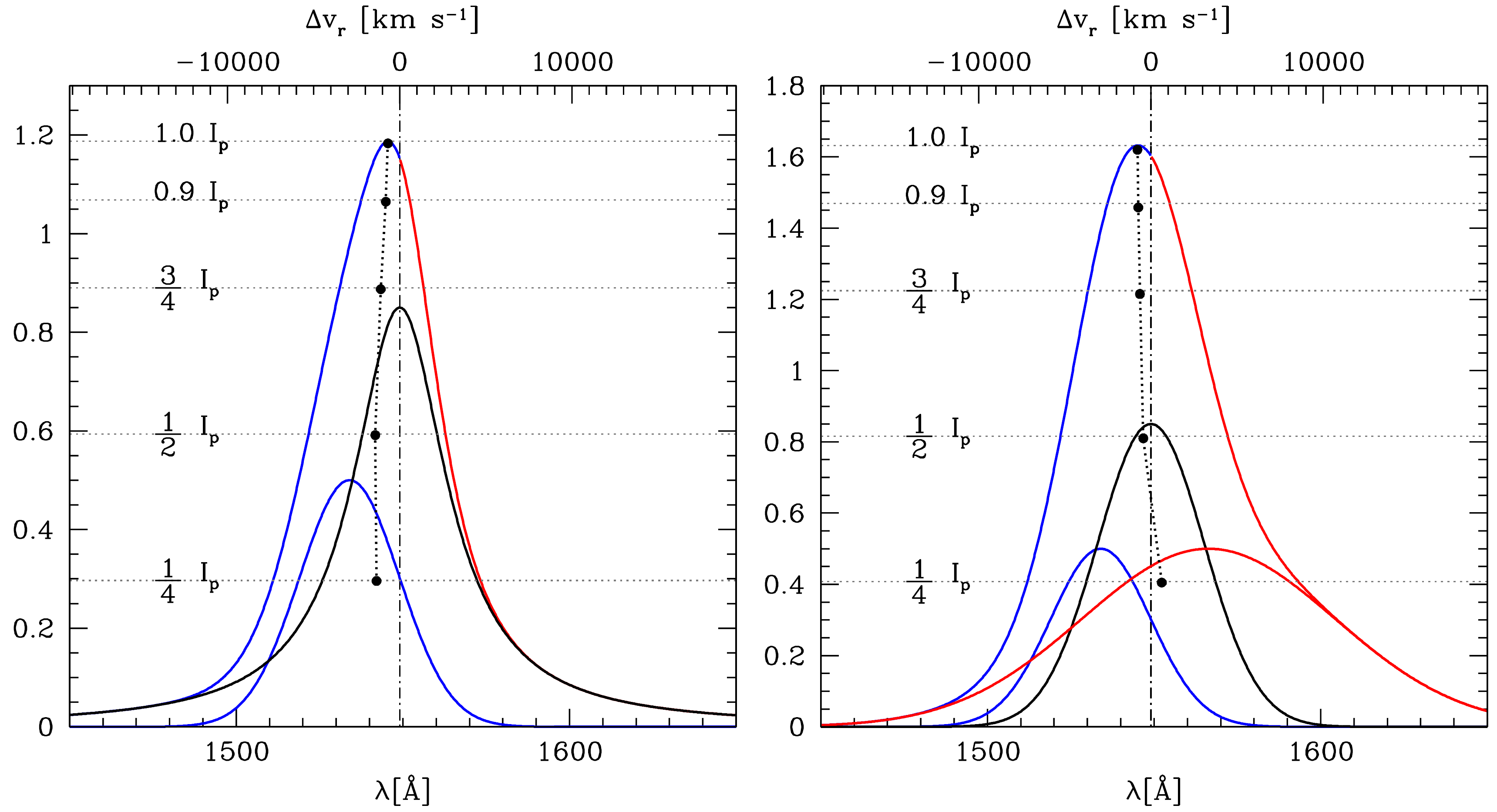} 
\end{minipage}
\hspace{0.25cm}
\begin{minipage}[t]{0.45\linewidth}
\centering 
\vspace{-1.8in}
\caption{ Interpretation of line profiles, for Pop A, and Pop. B (right). Fractional intensity levels where line centroids are measured are identified. Mock profiles are shown to represent the bare broad profile of any of the strongest emission lines of quasars. The left one is built on 2 components, as appropriate for Pop. A, which are the BC and a blue shifted excess BLUE. The blue shifted component is strong in HILs and weak in LILs.  Pop. B profiles are accounted for by three  components: in addition to the BC and BLUE, a redshifted VBC is needed to account for the prominent redward line profile asymmetry.   
\label{fig:3} }
\end{minipage}
\end{figure}
\subsection{Multi-component interpretation of emission lines}
\label{multi}

The broad profile of both LILs and HILs in each quasar spectrum can be modeled  by changing the relative intensity of three main components, as shown schematically in Fig. \ref{fig:3}.  

\begin{itemize}
\item  The broad component (BC), which has been referred to by various authors as the broad component,  the intermediate  component, or the central broad component \citep[e.g.,][]{brothertonetal94a,popovicetal02,adhikarietal16,kovacevicdojcinovicpopopvic15}. As mentioned, the BC is represented by a symmetric and unshifted profile  (Lorentzian for  Pop. A or Gaussian for Pop. B) and is therefore assumed to be associated with a virialized BLR subsystem. The { virialized} BLR could be defined as the subregion that is in the condition to produce FeII emission. 
\item The blue shifted component (BLUE).  A strong  blue excess in Pop. A \civonly\ profiles is not in doubt.  In some \civonly\ profiles like  the extreme Population A  prototype I Zw1 the blue excess   is by far the dominant contributor to the total emission line flux \citep{marzianietal96,leighlymoore04}.  BLUE is modeled  by one or more skew-normal distributions   \citep{azzaliniregoli12}. The ``asymmetric Gaussian'' use is, at present, motivated empirically by the often irregular appearance of the blueshifted excess in \civonly\ and \mgii.  
\item The very broad component (VBC). The  VBC was postulated because of typical \hb\ profile of Pop. B sources, that can be (empirically) modeled with amazing fidelity { (i.e., with no significant residuals above noise in the fit composite spectra)},  using the sum of two Gaussians, one narrower and almost unshifted (the BC) and one broader showing a significant redshift $\sim 2000$ km/s\ \citep[][]{willsetal93,brothertonetal94,zamfiretal10}. We expect a prominent CIV\ VBC associated with high ionization gas  in the innermost BLR \citep{sneddengaskell07,marzianietal10,wangli11,goadkorista14}.  {Past works indeed provided evidence of a VBC in \civonly\ and Balmer  lines \citep[e.g.,][]{marzianietal96,sulenticetal00c,punsly10,marzianietal10}.} Mirroring  the  definition of BLR, a VBLR may be defined as the region that is not able to emit significant FeII. Imposing significant FeII VBC emission in the multicomponent maximum likelihood fits of Population B H$\beta$ produces unrealistic emission patterns in the FeII blend. Agreement is restored only when VBC emission is assumed negligible.  
At high $L$, BLUE can  dominate in Pop. B \civ\ profile, too, but the VBC remains well-visible, especially in \hb.  
\end{itemize}

\subsection{LIL- and HIL-emitting regions}
\label{hillil}

The symmetric BC and the BLUE are consistent with a two-region model  proposed since the late 1980s  \citep{collinsouffrinetal88}, where the BC is emitted in a flattened distribution of gas while BLUE is associated with an high ionization outflow. A two region model such as a disk + radiation dominated wind is also qualitatively consistent with the data \citep{elvis00}. 

Considering intensity ratios using the full line profiles can be misleading and may yield to problematic inferences.  For example, the explanation of the Ly$\alpha$/\hb\ ratio has been one of the most challenging problems in the interpretation of line formation within the BLR \citep{netzeretal95}. It is easier to account for the observed ratios if BLUE and BC are kept separated. In the BLUE case, the ratio Ly$\alpha$/\hb\  is very high, consistent with relatively low density and high ionization ($ \approx 30$ close to the case A in the low-density limit, \citealt{osterbrockferland06}), while in the BC case Ly$\alpha$/\hb\ is low ($\approx 5 -  10$) which requires high density and low-ionization, following CLOUDY simulations \citep{marzianietal10}. It has proven possible to reproduce the profiles of the strongest broad emission lines along the MS changing the  relative proportion of the three components, BLUE, BC and VBC,  but assuming consistent component shift, width and shape parameter for all lines.  As a second example, it is interesting to consider the case of the \civ\ profile in broad Pop. B RL and RQ sources. The centroid shifts to the blue are modest for { both RQ and RL}; however, redshifted $c(\frac{1}{2})$ are not unfrequent among RL { but very rare among RQ
\citep{sulenticetal07,richardsetal11}}. Comparing the \civonly\ profile to \hb\ offers a clue: \hb\ can be accounted for by two Gaussians, one of them redshifted by $\sim 1000 - 2000$ km/s. With respect to the \hb\ profile, the \civonly\ lines shows a blueshifted excess (the BLUE component) that, even if not as strong as in Pop. A, has nonetheless the effect to symmetrize their profile. In some RL it may be completely absent, so that   centroid measurements of \civonly\ and \hb\  agree \citep{marzianietal96}. This  analysis allows us to infer two interesting facts: whenever BLUE is present we have CIV/\hb $\gg 1$\ (if BLUE is not detected in  \hb, a lower limit to CIV/\hb\ is measured from the noise); BLUE is apparently more prominent in Pop. B RQ than in RL, an effect that may be associated with the effect of the jet lateral expansion on the accretion disk wind \citep{sulenticetal15}.
 
The relative prominence of the three components can be accounted for in terms of balance   of radiation and gravitation forces, at least in RQ quasars.  If line emitting gas is optically thick to the Lyman continuum then the radiation will exert an outward acceleration that  is inversely proportional to the column density $N_\mathrm{c}$\ and directly proportional to the ionizing luminosity. The ratio between the radiative and gravitational  acceleration  for gas optically thick to the Lyman continuum  is: $r_\mathrm{a} = {a_{\mathrm{rad} } }/{a_{\mathrm{grav} } } \approx 0.176 \kappa L_{44} M_\mathrm{BH,8}^{-1} N_\mathrm{c,23}^{-1}$ where \mbh\ is  in units of 10$^8$ solar masses, and $L_{44}$ the bolometric luminosity in units of 10$^{44}$\ erg/s, $\kappa$ the fraction of the bolometric continuum that is ionizing HI  ($\lambda < 912$\AA), $\alpha \approx 0.5$\ \citep{marzianietal10}. The equation can take the form $r_\mathrm{a}   \approx 7.2\, {L}/{L_{\mathrm{Edd}}}  N_\mathrm{c,23}^{-1}, $ \ and  shows that the net outward acceleration is proportional to \lledd, and inversely proportional to column density $N_\mathrm{c}$.  If $r_\mathrm{a} \gg1$,  radiative acceleration dominates, as it is apparently the case of the high ionization gas emitting the blueshifted excess (i.e., BLUE). If $r_\mathrm{a} \ll1$, the emitting gas may be infalling toward the central black hole, yielding the observed redshifted VBC. This interpretation  is compatible with large $N_\mathrm{c}$\ values for the VBLR and would naturally explain why the VBC is observed in objects with low Eddington ratio, \lledd $\lesssim 0.1$. Similar considerations have been made  to explain the origin of an FeII redshift with respect to the broad FeII emission \citep{ferlandetal09,huetal12}. While the reality of FeII emission shifts is debatable \citep{sulenticetal12}, an infall scenario may well apply to inner BLR yielding a redshifted VBC, provided that column density is high enough to withstand radiation forces.

\section{Connection to accretion  parameters} 
\label{sec:accpar}

Several physical parameters (black hole mass \mbh, Eddington ratio, spin,  and the aspect angle $\theta$) are expected to affect the parameters of the E1 MS, even if in an indirect way as in the case of the spin. Establishing a connection between  a physical    and an observational set of parameters is precisely  the aim of the 4DE1 parameter space. Table 2 lists the main physical parameters on which the empirical 4DE1 parameters are expected to be dependent. We still have a very incomplete  view of the physics along the Eigenvector 1 sequence because we are able to make only coarse estimates of physical parameters. 
However, Eddington ratio and viewing angle $\theta$ are likely to be the main culprits affecting the location of a quasar along the MS, as discussed in the next sections. 

\subsection{Black hole mass}

The virial black hole mass can be written as $M_\mathrm{BH} = {f} r_{BLR}\delta v^{2}/G$,where ${f}$\ is the emitting region structure or form factor, $\delta v$ is a VBE (typically the line FWHM or its dispersion), and $r_\mathrm{BLR}$ is a characteristic distance from the black hole of the line emitting gas (i.e., in practice the distance derived from reverberation mapping or from the photoionization method).   There are two fundamental aspects to consider: evaluating \mbh\ requires consistent estimates of  $r_\mathrm{BLR}$ and $\delta v$, for different lines; in addition, the parameters entering the virial expression of \mbh\ (${f}, r_{BLR}, \delta v$) {\em all depend on the location of the E1 sequence.} It is an unfortunate circumstance that, to-date, this fact has not been taken into account in an exhaustive way. In the following  we just mention how the applications of the E1 trends on the computation of \mbh\ can improve the accuracy of its estimates.  
 
\subsubsection{The virial broadening estimator (VBE)}
 
There is a growing consensus that the line width of the strongest LILs (\hb\ and Mg$\lambda$2800) provide a reliable VBE, up to the highest luminosities \citep[e.g., ][]{trakhtenbrotnetzer12,mejia-restrepoetal16,sulenticetal17}.  UV intermediate emission lines -- have been found suitable as well, at least for low-$z$ quasars \citep{negreteetal13}, 
The HIL \civonly\ has been considered as a sort of taboo, as it offers a highly biased VBE \citep{baskinlaor05b,sulenticetal07,netzeretal07}.  The broadening of \civ\ is affected by shear velocity in an outflow, and its FWHM is not immediately offering a reliable VBE \citep[as recently discussed in the review of ][]{marzianietal17d}. Figure 1 of \citet{marzianietal17d} shows the bias  introduced into  \civonly\ \mbh\ computations along the MS by using uncorrected FWHM \civonly\ values. Actually, in Pop. A errors on \mbh\ as large as factor of 100 are possible if the \civonly\ FWHM is measured on the full \civonly\ profile. Even if corrections can be derived, whether the  \civonly\ width may be a reliable VBE remains controversial \citep[][and references therein]{denneyetal13,coatmanetal17,marzianietal17d}.  
It is possible to identify corrections that would reduce the scatter between \hb\ and CIV-based \mbh\ estimates to $\approx 0.33$ \citep{coatmanetal17,marzianietal17d}. In other words, applying corrections to the  \civonly\ FWHM, the \hb\ and \civonly\ FWHM lines would become equivalent VBEs. While important conceptually, these corrections may be cumbersome to apply in practice: since they are based on measures of the \civonly\ line shift (the nonparametric measure of \citealt{coatmanetal17} is  equivalent to $c(\frac{1}{2})$), they still require the knowledge of the quasar rest frame {  which is not always straightforward to estimate (as pointed out in Section \ref{intro}; see also  \citealt{hewettwild10} for a discussion of $z$-dependent biases)}. In addition, corrections are different for Pop. A and B, and considering the Eddington ratio bias implicit in flux limited samples \citep{sulenticetal15}, they may remain sample-dependent. 

We are now able to analyze systematic trends for the virial broadening of low-ionization lines along the E1 MS leaving aside random orientation effects that are expected to influence $\delta v$\ estimates. We can define a parameter $\xi$ yielding a correction to the observed profile: $\xi = {\mathrm{FWHM}_\mathrm{VBE}}/{\mathrm{FWHM}_\mathrm{obs}} \approx  {\mathrm{FWHM}_\mathrm{BC}}/{\mathrm{FWHM}_\mathrm{obs}}$, where the VBE  FWHM can be considered best estimated by the FWHM of the broad component of any line, FWHM$_\mathrm{BC}$. The  non-virial broadening affecting the integrated   profiles  of LILs along the E1 sequence is  due to different mechanisms (as mentioned in \S \ref{birdseye}): the A3/A4 spectral types are affected by outflow, while the \hb\ redward asymmetry may suggest an infall of the VBC emitting gas. However, the correction factor is modest  ($0.75 \le \xi \le 1.0$) for both \hb\ and \mgii, with $\xi \approx 1.0$\ in spectral types A1 and A2. In other words, a simple correction is sufficient to extract a VBE from the observed FWHM.  The correction can be evaluated for each spectral type or by applying individual source corrections as described in \citet{marzianietal17d}. 

\subsubsection{Estimates of $r_\mathrm{BLR}$}

We can distinguish   primary and secondary estimates of the radius of the BLR $r_\mathrm{BLR}$. Primary determinations come from reverberation mapping monitoring \citep[][and references therein]{petersonetal04a,peterson14} and  are measured from the time lag $\tau$\ yielded by  the peak or centroid of the cross-correlation  function between   continuum and   line light curves. Primary estimates can be also obtained from rest-frame, single-epoch UV spectra using the so-called photoionization method, as summarized below.  Secondary determinations are computed using the correlation between $r_\mathrm{BLR}$\ and luminosity that has been derived from reverberation-mapped sources  \citep[e.g.,][]{kaspietal00,kaspietal07,bentzetal09,duetal16}:   $r_\mathrm{BLR}  \propto L^a$, a$\approx$0.5 -- 0.65 \citep{kaspietal00,bentzetal06}.   

There are several caveats underlying the RM measure of $r_\mathrm{BLR}$\ and  its scaling law with luminosity   \citep{marzianisulentic12}.  This correlation has a considerable scatter \citep{marzianisulentic12} that is propagating itself on the mass scaling laws written in the form $M_\mathrm{BH} = M_\mathrm{BH}(L, FWHM) = kL^{a} FWHM^{b}$ \citep[e.g., ][]{vestergaardpeterson06,trakhtenbrotnetzer12,shenliu12}.  In addition, the $r_\mathrm{BLR}$\ -- $L$ scaling relation depends on dimensionless accretion rate \citep{duetal16}. Therefore the scaling laws should be redefined for at least Pop. A, extreme Pop. A (as actually done by \citealt{duetal16}) and Pop. B along the E1 sequence. 


The ionization parameter $U$ can be written as $U = {Q(H)}/{4 \pi r_\mathrm{BLR}^2 n_\mathrm{H} c}  \propto  {L}/{r^{2}n}$\ \citep{netzer13}, where $Q(H)$ is the number of hydrogen ionizing photons.  The radius of the BLR then can be recovered as $r_\mathrm{BLR} = \frac 1{h^{1/2} c} (n_\mathrm{H} U)^{-1/2} \left( \int_{0}^{\lambda_{Ly}} f_\lambda \lambda d\lambda \right) ^{1/2} d_\mathrm{C} $, where $d_\mathrm{C} $  is the comoving distance to the source \citep{hoggfruchter99}, and the integration is carried out to the Lyman limit $\lambda_\mathrm{Ly} = 912 $ \AA.  Clearly, the ionizing photon flux $U$\nh\ is a measurement of the exposure of the BLR to ionizing photons and hence has an intrinsic, strong dependence  on $r_\mathrm{BLR}$.  The first explorative estimates of \mbh\ using the photoionization method  were based on the rough similarity of AGN spectra, and on the consequent assumption of  constant  $U$\ or of constant product $U$\nh\ \citep{padovanirafanelli88,wandeletal99}.  The consideration of UV line ratios which can be used as   diagnostics constraining ionizing photon flux, and, in some cases, \nh\ \citep{verneretal04,negreteetal13,negreteetal14a} yielded a major improvement. It is remarkable that the  $r_\mathrm{BLR}$ estimates from the $U$\nh\ product   and the H$\beta$\ $r_\mathrm{BLR} = c \tau$  from reverberation mapping are in very good agreement, at least for 12 low-$L$ AGNs \citep{negreteetal13}.   In both cases (photoionization and reverberation) we are trying to give {\em one}\ number that should be representative of a very complex region, probably stratified, perhaps chaotic. Evidently, the 1900 blend lines { of \aliii\ and \siiii\ (but not \ciii!)} trace the physical conditions of the  H$\beta$\ emitting gas with maximum responsivity, { which is not surprising since  \aliii\ and \siii\ are lines emitted by ionic species with low-to-intermediate ionization potential, $\approx 15 - 20$ eV \citep{negreteetal12}.  }


A second remarkable aspect  emerges from the analysis of the 1900 blend. \citet{negreteetal13}  showed that the solutions yielding the $U$\nh\  product differ significantly if the \ciii\ is involved with other high ionization lines, or if \civ, \aliii, \siiii, \siiv, \siii\ are considered. Their Fig. 2 clearly shows  that \ciii\ involves a solution of lower density and higher ionization. This can be interpreted as evidence of stratification within the BLR. If extreme Population A sources are considered, \ciii\ is weak in their spectra and only the   low-ionization solution remains. { The \ciii/\siiii\ ratio may be further lowered because of the relatively soft spectral energy distribution of xA sources \citep{casebeeretal06}.} Apparently, the xA sources have their LIL/IIL emission dominated by dense, low-ionization gas (as  lower density and column density gas is being pushed away from the BLR in a high ionization outflow with \civ/\ciii$\gg$1). 

The application of the photoionization method has been extended to high-$L$ quasars where it remains basically untested: apart from the absence of systematic effects with scaling laws prediction, the precision and accuracy of individual photoionization estimates have nothing they can be compared with. \citet{negreteetal13} suggested a tentative correction on the basis of the equivalent width ratio between AlIII and CIII] (an E1 correlate) but the correction is highly uncertain. More objects are needed to better understand the behavior of the photoionization-derived $r_\mathrm{BLR}$ along the E1 sequence.  In principle, however, the photoionization method has the potential to reduce the intrinsic scatter in \mbh\ determination, if it is really able to produce $r_\mathrm{BLR}$\ estimates in close agreement with the RM $c\tau$.  

\subsection{Orientation effects}

There is no doubt that orientation effects influence the FWHM of \hb\ in radio loud type-1  quasars (as mentioned in Sect. \ref{sec:unif}): the line is systematically broader in FRII than in CD sources \citep[e.g.,][]{willsbrowne86,rokakietal03,sulenticetal03}. It is reasonable to assume that  effects of similar amplitude are present also in RQ sources \citep{marzianietal01} although defining a reliable orientation indicator has proved elusive. Recent results suggest that  orientation is affecting the shifts of [OIII] (as well as the EW, \citealt{risalitietal11}) and of FeII:  face-on sources should show no FeII shifts and high amplitude blueshifts; more inclined sources should show FeII redshifts (associated with an equatorial inflow) and no net [OIII] blueshifts \citep{huetal08,boroson11}. While [OIII] shifts and EWs are most likely affected by orientation, orientation does not appear to be the main parameter, if [OIII] emission is considered in different parts of the E1 sequence. Highly blueshifted sources (such as the BOs) are apparently associated exclusively with high Eddington ratio, more than those with a face-on orientation \citep{marzianietal03b,xuetal12}. The distribution of [OIII] blueshift amplitudes along the E1 sequence is qualitatively not consistent with orientation-only effects. As already mentioned, the reality of FeII high-amplitude  redshifts  that would serve as indicators has been questioned recently \citep{sulenticetal12}. Even if we have not yet identified an orientation indicator, orientation effects most likely account  for a large part of the scatter in \mbh\ determinations.{   The observed velocity can be parameterized as  $  v^{2}_\mathrm{obs} = v_\mathrm{iso}^{2}/3 + v_\mathrm{Kepl}^{2} \sin^{2}\theta$, and if $v_\mathrm{iso}/v_\mathrm{Kepl}\approx 0.1$, where  $v_\mathrm{iso}$\ is an isotropic velocity component,  and $v_\mathrm{Kepl}$\ the Keplerian velocity.  For a  geometrically thin disk, it implies} $v_\mathrm{obs} \approx v_\mathrm{Kepl}/\sin\theta$ {( if the FWHM is taken as the $v_{\rm obs}$, and $v_{\rm Kepl} = 0$\ i.e., in the case of isotropic velocity dispersion,  $f  = \frac{3}{4}$)}.  If the VBE estimates are not corrected beforehand for orientation,  the structure (or form) factor is $f  \propto 1/\sin^{2}\theta$\ \citep{jarvismclure06}. { The structure factor is also expected to depend on physical parameters (Eddington ratio, metallicity, disk temperature etc.) apart from aspect effects. }

\subsection{The structure factor}

The  velocity resolved RM yields an amazing variety of velocity fields for different objects \citep[e.g.,][]{grieretal13,peterson17}.  The evidence favoring a rotational component has been steadily growing, in part due to deep, single-epoch spectropolarimetric observations \citep[e.g.,][]{smithetal05,afanasevetal15} that revealed the polarization vector changes expected from the Keplerian  velocity field as seen from an equatorial scatter. Systematic outflows have been made evident by the ubiquitous blueshifts of the  \civ\ emission line.   A hint to the BLR structure is provided by the FWHM/$\sigma$ ratio, where $\sigma$ is the velocity dispersion of the profile. For Pop. A, the ratio is low, while it is close to the value expected for a Gaussian in Pop. B \citep{collinetal06,kollatschnyzetzl11}. The implication drawn from this result by \citet{kollatschnyzetzl11} is that broader lines imply faster rotation, which is consistent with the inferences based on the 4DE1 context (\S \ref{acc}). The structures underlying the typical Pop. A and B broad profiles (\S \ref{acc}) are however still unclear.  

In Pop. A the Lorentzian profiles are consistent with an extended accretion disk seen at moderate inclinations \citep{dumontcollinsouffrin90e}, but  in the context of AGN, several velocity fields can produce Lorentzian-like profiles \citep[e.g., ][]{mathews93,netzermarziani10,czernyetal17}. Recent works suggest a disk-core with wings produced in a region of enhanced vertical velocity dispersion with respect to the disk \citep{goadetal12}. 

On the other hand  the quasi-symmetric profiles of the   Pop. B LILs imply virial/Keplerian motion with  stratified physical properties  \citep[e.g.,][]{fausnaughetal17}.  An intriguing possibility is that the VBC may be emitted by the inner accretion disk \citep{bonetal09a,storchi-bergmannetal17}, with the BC masking the double-peaked accretion disk profile expected if the disk external radius is not extremely large, $\gtrsim 10^{3} R_\mathrm{g}$.

\citet{collinetal06} derived different values for $f_\mathrm{S}$ for Pop. A and B, 2.1 and 0.5 respectively, with a substantial scatter.  The distinction between Pop. A and B therefore appears to be the minimal criterion to reduce scatter in \mbh\ estimates, {also}  because comparison of the same  line width measure is not easy to understand if profile shapes are different. { A more refined approach may consider individual spectral types along the quasar MS.}

\subsection{\lledd}

The \mbh\ scaling laws provide a simple recipe usable with single-epoch spectra of large samples of quasars. Estimates of the Eddington ratio are derived by applying a constant bolometric correction to the observed luminosity,  typically   a factor 10--13 from the flux at 5100 \AA\  and 3.5--5 from the measured at 1450 \AA\ \citep{elvisetal94,richardsetal06,krawczyketal13}. Bolometric corrections  most likely depend on the source location along the MS: { the anti correlation between UV luminosity and optical-to-X spectral index (between 2500 \AA\ and 2 KeV) may go in the sense of softer continua in Pop.  A  \citep{laoretal97b,steffenetal06}.}  In addition, bolometric corrections  are most likely luminosity as well as orientation dependent \citep{runnoeetal13}. However,  our group did not carry out a systematic study as yet, and   applied almost always  constant corrections.


\subsection{Why Does Ionization Level Decrease With Increasing \lledd?}
\label{ulm}

As early as in the 2000s, a puzzling but remarkable aspect appeared to be the decrease in ionization degree at high Eddington ratio: Pop. A  spectra show  strong FeII, low [OIII] and \civ, while Pop. B sources with \lledd $\lesssim 0.2$\ show prominent (high EW) HILs. 

The ionization parameter $U$ can be rewritten in terms of $L/M_\mathrm{BH}$ and \mbh, under several assumptions.   The number of ionizing photons is $Q(H)\approx  \kappa L  /<h\nu>  $.  A typical AGN continuum
as parameterized by \citet{mathewsferland87} yields $<\nu>\approx 1 \times
10^{16}$\ Hz and $\kappa \approx$ 0.5.  The velocity field for the LIL-emitting gas is  mainly virial, so that we can write: FWHM $ \propto M^{1/2}{r}^{-1/2}$.  We can consider that the ratio I(\aliii)/I(\siiii) is a density-dependent diagnostics (almost independent of the ionization parameter)\footnote{ While the ratio I(\siiii)/I(\ciii)\ may depend on the spectral energy distribution \citep{casebeeretal06}, the ratio I(\aliii)/I(\siiii) is unlikely to show a strong dependence because the involved ionic species have similar ionization potential, and the transition upper levels are close in energy above ground level.}, and that  $I({\rm  Al ~ III})/ I({\rm Si~  III]}\lambda 1892)$\ and  $ I({\rm Si~  III]}\lambda 1892)/I({\rm  C ~ III}])$ are inversely correlated with FWHM. Using the trend found by \citet{willsetal99},  $n \propto\ $FWHM$^{-4/3}$. At this point it is possible to write \rfe\ as a function of \lledd\ and \mbh\ using two approaches \citep{marzianietal01}. 

\begin{itemize}
\item Considering that $L/M_\mathrm{BH} \propto$ FWHM$^{-2}$, with \mbh\  estimated from X-ray variability, not from the virial relation, to avoid circularity.    The  expression for  $U \propto  \frac{L}{r^2 n} \propto (L/M_\mathrm{BH})^{-5/3} M^{-1}$\ follows from $n \propto\ $ FWHM$^{-4/3}$ and $r \propto M/$FWHM$^{2}$. 
\item Adopting the $ r \propto L^{a}$ scaling law, and  considering again that $n \propto $FWHM$^{-4/3}$\ and FWHM $ \propto M_{\rm BH}^{1/2}{r}^{-1/2}$,   $U \propto  \frac{L}{r^2 n} \propto \frac{L^{1-2a}}{n}  \propto  {(L/M_\mathrm{BH})^{1-\frac{8}{3}a}}{M_{\rm BH}^{\frac{5-8a}{3}   }}$. With $a=1$, the first expression is recovered. For $a = 0.5$\ there is a weaker dependence on ($L/M_\mathrm{BH}$) to { the power of}  $-\frac{1}{3}$.\footnote{Note that, in this case, it is not necessary that $a = 0.5$\ to recover spectral similarity. Different spectral types are likely to be associated with different scaling with luminosity. }
\end{itemize}
The grid shown in \citet{marzianietal01} obtained using the first approach does not make it possible to derive orientation angle $\theta$ and \lledd\ for individual quasars, as it was computed for a fixed \mbh.  These considerations  provide  a first account of why   sources at the high-$L/M_\mathrm{BH}$ tip  are  associated with a lower ionization degree. Other explanations are possible as well: the line emitting gas is shielded from the most intense UV radiation, for example by an optically and geometrically thick slim disk \citep{wangetal14a} or by an inner, over-ionized failed wind \citep{leighly04} but the intensity ratios of $I({\rm  Al ~ III})/ I({\rm Si~  III]}\lambda 1892)$\ and $ I({\rm Si~  III]}\lambda 1892)/I({\rm  C ~ III}])$\ point  toward an increase of the emissivity-weighted density values of the line emitting gas. This alone yields a decrease in $U$ at high \lledd, all other parameters left unchanged.

\begin{figure}[htp!]
\begin{minipage}[t]{0.45\linewidth}
\centering
\includegraphics[width=7.5cm]{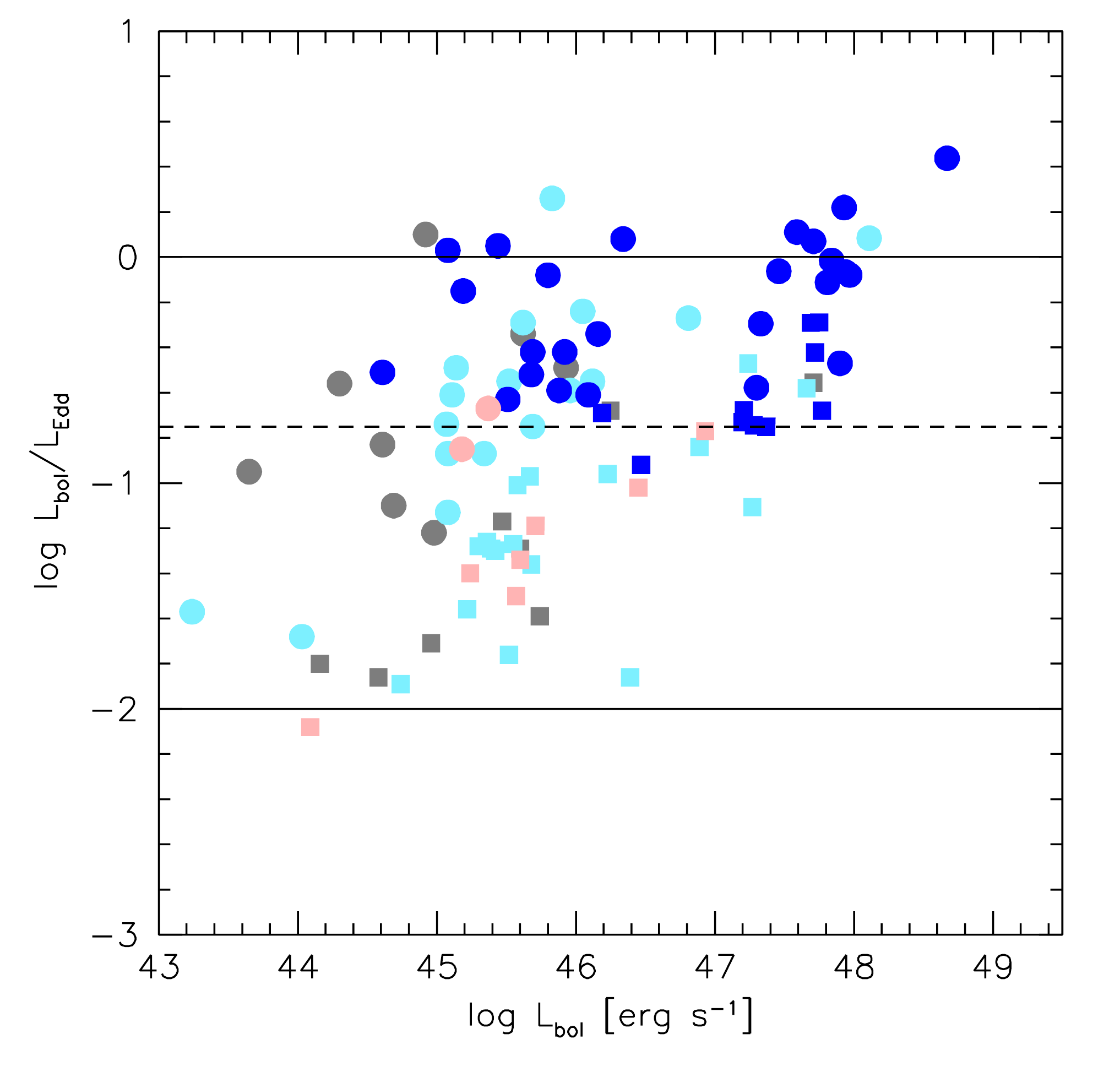} 
\end{minipage}
\hspace{0.25cm}
\begin{minipage}[t]{0.45\linewidth}
\centering 
\vspace{-2.75in}
\caption{Behaviour of \civonly\  shifts in the plane  Eddington ratio versus bolometric luminosity. Circles indicate Pop. A, squares Pop. B. Color coding is as follows: blue, $c(\frac{1}{2})  \le -1000$ km s$^{-1}$; pale blue: $ -1000  < c(\frac{1}{2}) \le -300$ km s$^{-1}$; grey $ -300  < c(\frac{1}{2})  \le  300$ km s$^{-1}$; pale red: $ 300  < c(\frac{1}{2}) \le 1000$ km s$^{-1}$; red $    c(\frac{1}{2}) > 1000$ km s$^{-1}$. The sample sources correspond to the Hamburg-ESO sample  (all above    $L \gtrsim 10^{47}$  erg s$^{-1}$) and FOS-based control sample of \citet{sulenticetal17}. \label{fig:mlmhe} }
\end{minipage}
\end{figure}

\begin{figure}[h!]
\begin{center}
\includegraphics[width=18.5cm]{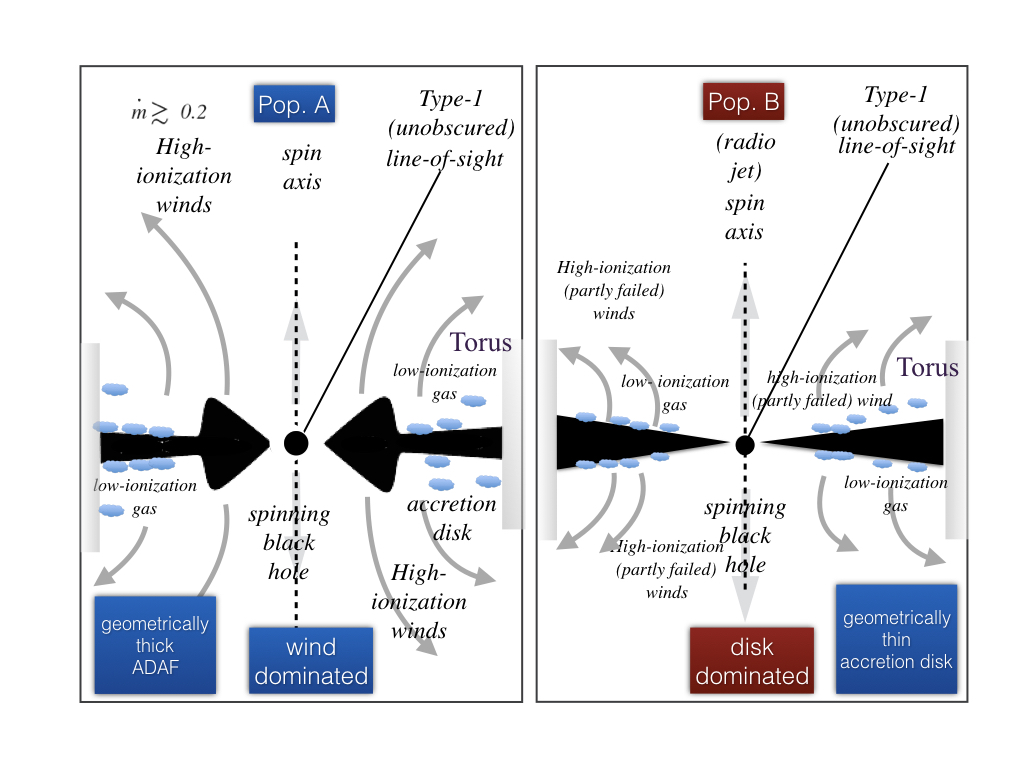}
\vspace{-.5in}
\end{center}
\caption{Different structure for Pop. A (left) and B (right), with a slim and a flat disk respectively. The sketch is not drawn to scale and the relation between line emitting regions (shown here as clouds) and accretion disk structure is still debated.  
See \S \ref{acc} for a more detailed explanation. }\label{fig:fm1} 
\end{figure}

\section{Self-similarity of the accretion process}
\label{sec:self}

The accretion process is apparently largely self-similar  over several order of magnitudes in black hole mass and luminosity. The general fundamental plane of accreting black holes emphasizes the invariance of the accretion disk-jet scaling phenomena \citep{merlonietal03}. The Eigenvector 1 scheme is also emphasizing a scale invariance, albeit more oriented toward the radiatively-efficient domain and, in the quasar context, limited to unobscured type-1 quasars.  As far as the radiatively efficient accretion mode is concerned, the invariance may hold over a factor of almost $\sim 10^{10}$ in solar mass \citep{zamanovmarziani02}. Restricting the attention to RQ quasars, very large blueshifts are observed over at least four orders of magnitude in luminosity. The distribution of data points in Fig. \ref{fig:mlmhe} is clearly affected by selection effects; however, it shows that large \civonly\ $c(\frac{1}{2}) \lesssim -1000$ km s$^{-1}$\ do occur also at relatively low luminosity. Outflow velocities are apparently more related to Eddington ratio than to $L$ or $M_\mathrm{BH}$: large $c(\frac{1}{2})$\  blueshifts may occur only above a threshold at about  \lledd $\approx 0.2 - 0.3$ which may in turn correspond  to a change in accretion disk structure \citep[][and references therein]{abramowiczstaub14}. Intriguingly, the threshold matches the FWHM(H$\beta$)$ \approx 4000$ km s$^{-1}$ limit separating Population A from Population B (if $ L  \lesssim 10^{47}$  erg s$^{-1}$). At this limiting FWHM we observe also a change in \hb\ profile shape, from Lorentzian-like to double Gaussian \citep{sulenticetal02,zamfiretal10}.

\subsection{Pop. A and B: a different accretion structure}
\label{acc}

We propose the two panels with the sketches shown in Fig. \ref{fig:fm1} as a pictorial view of quasars accreting at  high (left) and low rate respectively \citep[c.f.][]{marinuccietal12,marzianietal14,luoetal15}.  The main theoretical prediction is that we expect an inner accretion disk region assimilable to a slim disk  \citep{abramowiczetal88,szuszkiewiczetal96,franketal02}.  Apart from this,  the structure of the BLR and its relation to the accretion disk structure remains unclear.  

In Pop. A sources, as mentioned before (\S \ref{ulm}), it is tempting to speculate that LILs may be favored with respect to HILs by the shielding of the hottest continuum due to the slim disk geometry. However, the question remains whether the ionized outflow we see in \civ\ is associated with the central cone defined by the walls of the slim disk (which may be much steeper than the ones deducted in the cartoon, \citealt{sadowskietal14}). In Pop. A, the \civonly\ shifts are largest  but the \civonly\  EW is lowest (Pop. A includes weak lined quasars,  \citealt{diamond-stanicetal09,shemmeretal10}), which may imply that the gas is over-ionized or, alternatively, that the FUV continuum is absorbed/weakened, as in the case of emission from the shielded part of the disk between the slim structure and the torus { \citep[][]{luoetal15}}. Against the latter interpretation goes  the higher  ratio of radiation-to-gravitation forces of Pop. A which yields a higher terminal velocity and hence systematically large blueshifts in the HILs:  continuum seen by the gas should at least roughly correspond to the continuum seen by us. 

In the case of a flat-disk (Pop. B right-panel), the problem of disk wind over-ionization may be solved by the failed wind scenario \citep{murrayetal95}:   inner gas may offer an effective screen and only shielded gas is efficiently accelerated { \citep{leighly04}}. Models of disk-wind systems are successful in reproducing the profiles of Balmer lines \citep{flohicetal12}. To explain the redward asymmetries often observed in Pop. B, either Balmer lines  are emitted in an infall scenario (which require large column density to withstand radiation forces) or the accretion  disk itself could be emitter. In the latter case the  redward asymmetry   could be ascribed in full to transverse and gravitational redshift \citep{bonetal09a,bonetal15}. 

Apart from these considerations, the sketch of  Fig. \ref{fig:fm1} raises more questions than it provides answers.  For example, can the inner part of the torus contribute to the velocity dispersion and yield a Lorentzian profile \citep{goadetal12}? This question  raises a conflict with the virial assumption for Pop. A sources, whereas the line wings are expected to be emitted closer to the black hole. Electron scattering may also produce extended line wings in the Balmer lines \citep{laor06}.  Roles of  magnetic fields and of black hole spin are not   considered although presumably important, black hole spin because of its effect on the inner accretion disk temperature, and magnetic fields because they may provide an acceleration mechanism for disk wind \citep{emmeringetal92}.

\subsection{The quasar main sequence at high luminosity}
\label{highl}

We still lack a comprehensive view of the MS at high $L$\ (we consider high-luminosity sources the quasars with bolometric $\log L \gtrsim 47$ [erg/s]), not last because the \hb\ spectral range is accessible only with IR spectrometers, and high-luminosity quasars are exceedingly rare at $z \lesssim 1$. The main effect that we expect in the optical plane of the 4DE1 space is related to their systematic increase in mass:  if the motion in the LIL-BLR is predominantly virial ($M_\mathrm{BH} \propto r \delta FWHM^{2}$) and the BLR radius follows a scaling power-law with luminosity ($r \propto L^\mathrm{a}$), then FWHM $\propto (L/M_\mathrm{BH})^{-1} L^\frac{1-a}{2}$. Assuming that \lledd\ saturates toward values  $ \mathcal{O}({1})$ \ \citep{mineshigeetal00},    the {\em minimum} FWHM should show a clear trend with luminosity. This prediction has been confirmed  by joining samples covering more than 4dex in $\log L$, from 44 to 48 \citep{marzianietal09}. At each $L$ there is a large spread of value that reflect the dependence of FWHM\ by Eddington ratio and mass.  

\begin{figure}[htp!]
\begin{minipage}[t]{0.5\linewidth}
\centering
\includegraphics[width=3.5in]{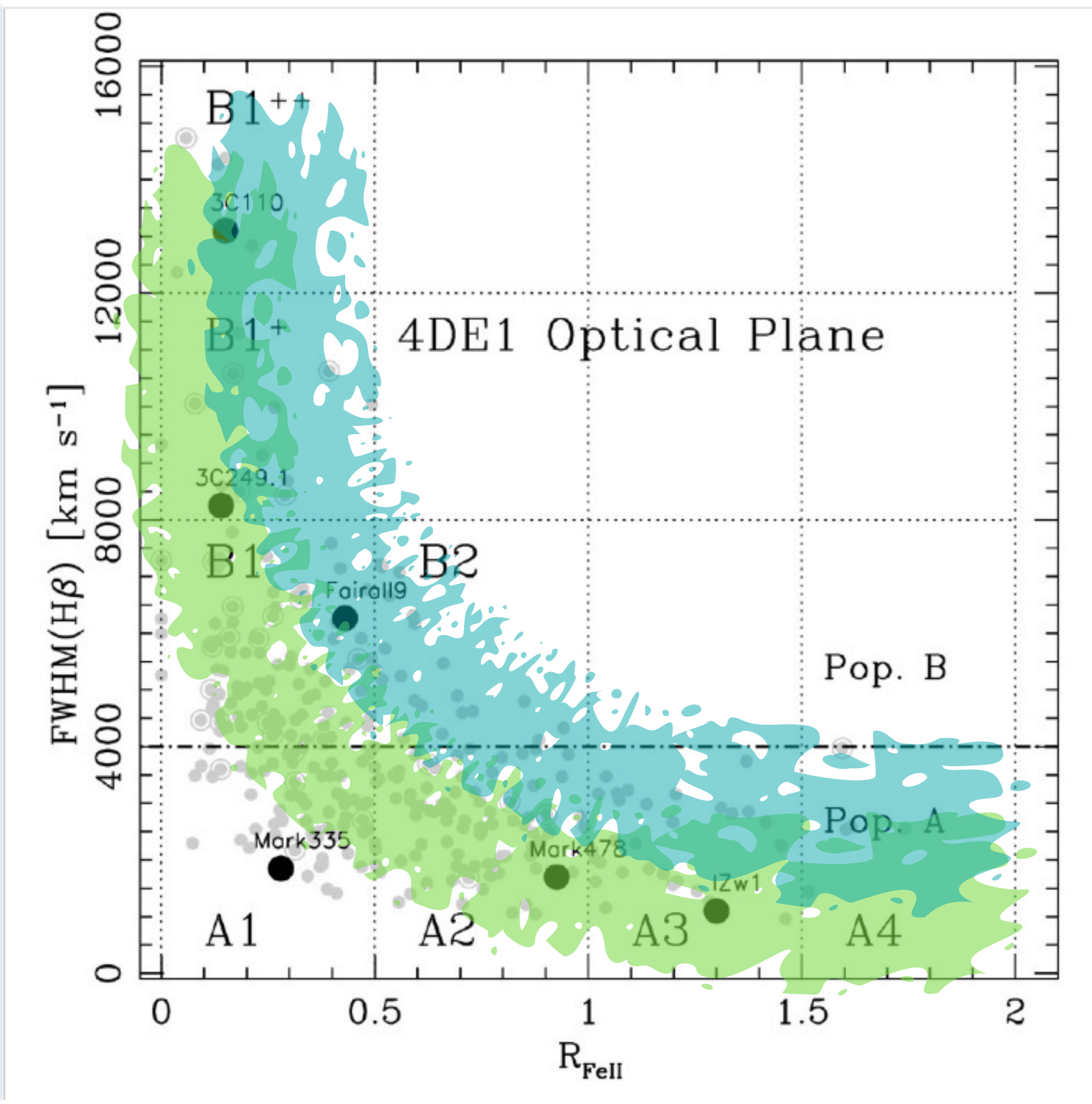}
\end{minipage}
\hspace{0.5cm}
\begin{minipage}[t]{0.5\linewidth}
\centering 
\vspace{-3.5in}
\caption{ Schematic representation of the E1 MS displacement in high luminosity samples.  The dark green shaded area represents a high luminosity $\log L \gtrsim 47$\ sample. In that case, NLSy1s are not anymore possible, if the virial assumption  holds up to the Eddington limit. }
\label{fig:mshighl}
\end{minipage}
\end{figure}

At high $L$, the MS becomes displaced toward higher FWHM values in the OP of the E1, as schematically represented in the diagram of Fig. \ref{fig:mshighl}. At present, we cannot say whether there is also a luminosity  effect on FeII prominence: relatively few sources are available to map the broad distribution of \rfe. Reports of an FeII Baldwin effect    appear unconvincing    and contradicted by the best available observations of the \hb\ range at high luminosity.  In addition, observations of high-$L$\ quasars from flux-limited surveys are subject to a strong Eddington ratio bias \citep{sulenticetal14} which may instead suggest an anti-Baldwin effect in FeII \citep{kovacevicetal10}.  The Eddington ratio bias involves the preferential selection of higher \lledd\ sources at higher $z$. Since \rfe\ strongly depends on \lledd\ (\S \ref{ulm}), low \rfe\ might be preferentially lost at high-$z$.

The comparison between low- and high-ionization lines has provided insightful constraints on the BLR at low-$z$\ \citep{marzianietal96}. The comparison of  \hb\ and \civ\ is yielding interesting clues also at high $L$ \ \citep{sulenticetal17,bisognietal17,shen16}.  Perhaps the most remarkable fact is that a LIL-BLR appears to remain basically virialized \citep{marzianietal09,sulenticetal17}. The \civ\ blueshift depends on luminosity   (the median $c(\frac{1}{2})$ is $\approx$\ 2600 \kms\ and 1800 \kms\ for Pop. A and B RQ at high-$L$ against less than 1000 \kms at low-$L$), but the dependence is not strong, and can be accounted for in the framework of a radiation driven winds. Assuming \lledd $\approx$ const. (as  is the case if a strong Eddington bias yields to a narrow range of \lledd\, as in high-$z$, present-day flux-limited samples), a luminosity dependence for $c(\frac{1}{2})$\ arises in the form $ \propto L^{1/4}$. If $L$ is restricted to a narrow range, the \lledd\ dependence dominates. If \lledd\ and $L$\ both span significant ranges, a multivariate analysis confirms the concomitant dependence on both parameters \citep{sulenticetal17}.
 
 

\section{Extreme Population  A sources: extreme radiators and potential distance indicators}
\label{sec:xa} 
 
Sources belonging to spectral types A3 and A4 (i.e., satisfying the criterion \rfe$>1$)\ are found to be   radiating at the highest Eddington ratio. They show a relatively small dispersion along a well-defined, extreme value { of order $\mathcal(1)$}\footnote{The exact values depend on the normalization assumed for \mbh\ estimates and for the bolometric correction.}  \citep{marzianisulentic14}. The xA selection criterion is consistent with  the parameter $\propto $ \ FWHM/$\sigma + $\rfe   used to identify  super-massive extremely accreting BHs \citep[SEAMBHs][]{wangetal13,duetal16}. Since xA sources show Lorentzian profiles, the criterion based on \rfe\ should be sufficient unless relatively broad profiles  with FWHM$\gtrsim$4000 \kms\ are considered, a case still under scrutiny.   

If $L/M_\mathrm{BH} \approx const.$, then the luminosity can be retrieved once the mass is known. xA sources show very similar spectra over a broad range of luminosity. The self-similarity in terms of diagnostic line ratios justifies the use of the scaling law $r_\mathrm{BLR} \propto L^{0.5}$\ that implies spectral invariance. The luminosity can then be written as  $L \propto (\delta v)^{4}$, where $\delta v$ is a suitable VBE. It is interesting to note that this relation is in the same form of the Faber-Jackson law as  originally defined \citep{faberjackson76}. Assuming that spheroidal galaxies are virialized systems, radiating at a constant $L/M_\mathrm{BH}$\ ratio implies that  their luminosity is $L \propto \sigma^{4}/\mu_{B}$, where $\mu_{B}$ is the  surface brightness within the effective radius. Later developments yielded a different exponent for the dependence on $\sigma$\ \citep[e.g.,][and references theerin]{donofrioetal17}, as the assumption of a constant surface brightness proved untenable. There should be no similar difficulties for quasars, although statistical and systematic sources of error should be carefully assessed \citep{marzianisulentic14}. The distance modulus computed from the virial equation,	$\mu  = 2.5 [\log  L(\delta v) - BC] - 2.5 \log  (f_{\lambda} \lambda) - const. + 5 \cdot \log (1+z)$, where $f_{\lambda} \lambda$\ is the flux at 5100 \AA, and BC the bolometric correction,
is in  agreement with the expectation of the concordance $\Lambda$CDM cosmology, providing a proof of the conceptual validity of the virial luminosity $L \propto (\delta v)^{4}$ \ estimates \citep[a plot of $\mu$ for several quasar samples is provided by][and by \citealt{negreteetal17}]{marzianietal17}.

\section{Conclusion}

The quasar MS provides a tool to systematically organize quasar multifrequency properties. It allows the identification of  spectral types with fairly well-defined spectral properties.  { In this paper, we have  described a simple parameterization that is able to describe the quasar emission line profiles, as well as a heuristic technique motivated by the internal line shifts yielding the separation of  components  in different physical conditions.

We then  considered the MS defined in the so-called optical plane of the 4D eigenvector 1 parameter space, and analyzed several correlates involving the profile of  \hb\ and \mgii\ (the representative LILs), of the \civ\   (the representative HIL), broad-line UV diagnostic ratios that provide trends in density and ionization level, \oiiiopt\ shifts, the prevalence of radio-loudness, and the soft X-ray spectral slope. 

The trends defined in this paper offer a consistent, empirical systematization of quasar properties for $z \lesssim 1$, low-to moderate luminosity quasars. The difference between Pop. A and B is  evident at the extremes: spectral type A4 sources typically show very strong \feii\ (\rfe $\gtrsim 1.5$), blueward asymmetry in \hb, large \civ\ blueshift, weak and blue shifted \oiiiopt\ \citep{negreteetal18}. Extreme Pop. B show undetectable \feii, very broad red-ward asymmetric \hb\ profiles, small-amplitude \civ\ shifts, prominent \oiiiopt. There is evidence that the two populations represent structurally different objects. The blue shifted excess (BLUE) is interpreted as due to outflowing gas dominating HIL emission only in Pop. A (unless sources at very high $L$\ are considered). A redshifted VBC is present only in Pop. B, for values of FWHM \hb\ above 4000 \kms, and has been interpreted as  due to gas close to the central continuum source.}  The relative balance between gravitational and radiation forces (i.e., \lledd) appears as a major factor influencing {\em both} the dynamics and the physical conditions of the line emitting gas (Sect. \ref{highl}), and an accretion mode change may be associated with a critical \lledd\ $\sim 0.2 \pm 0.1$, leading to the  two quasar populations: A (wind-dominated, following \citealt{richardsetal11}), and B (virial or disk-dominated). 

Luminosity trends are weak as they become significant only over a wide luminosity range   $\sim 10^{43} - 10^{48}$ erg/s. They involve an overall increase in virial line broadening (LILs) and an increase of blueshift frequency and amplitude consistent with the dominance of radiation forces (HILs).  

{ The presence of a virialized subregion identified along the MS at low-$L$ as well as at high-$L$ has important consequences.}  xA quasars at the high \rfe\ end (spectral type A3 and A4)  may be suitable as Eddington standard candles since their Eddington ratio scatters around a well-defined value \citep[e.g.,][]{negreteetal17}.

The contextualization offered by the MS is instrumental to the development of better-focused physical models along the quasar main sequence.  As  examples of the advantages offered by the E1 MS, we just mention the possibility of contextualizing orientation effects in RQ quasars, and of performing a meaningful comparison between RL and RQ samples with similar optical properties.  An aspect still  to  clarify is  the connection between disk structure and the dynamics of the line emitting gas. 


\section*{Author Contributions}

PM wrote the paper, with suggestions and comments from most of the authors who contributed to the activity of the research group.

\section*{Funding}

A.d.O., and M.L.M.A. acknowledge financial support from the Spanish Ministry for Economy and Competitiveness through grants AYA2013-42227-P and AYA2016-76682-C3-1-P. J. P. acknowledges financial support from the Spanish Ministry for Economy and Competitiveness through grants AYA2013- 40609-P and AYA2016-76682-C3-3-P. D. D. and A. N. acknowledge support from grants PAPIIT108716, UNAM, and CONACyT221398. E.B. and N. B. acknowledge grants 176003 ''Gravitation and the large scale structure of the Universe'' and 176001 ''Astrophysical spectroscopy of extragalactic objects'' supported by the Ministry of Education and Science of the Republic of Serbia. 

\section*{Acknowledgments}

It is a  pleasure to thank the SOC of the meeting  Quasar at all cosmic epoch for inviting a talk on the quasar main sequence. 


\bibliographystyle{frontiersinSCNS_ENG_HUMS} 







\end{document}